\documentclass[12pt]{iopart}
\usepackage{iopams}  
\usepackage{lscape}
\usepackage{graphicx}
\usepackage{threeparttable}
\usepackage{multirow}
\usepackage{color}
\newcommand{\mnras}{MNRAS}
\newcommand{\apjs}{ApJS}
\newcommand{\apj}{ApJ}
\newcommand{\apjl}{ApJL}
\newcommand{\prd}{Phys.~Rev.~D}
\newcommand{\sovast}{Soviet Ast.}
\newcommand{\nat}{Nature}

\begin{document}

\title[New tests of LLI of gravity with binary pulsars]{New tests
  of local Lorentz invariance of gravity with small-eccentricity
  binary pulsars}

\author{Lijing Shao$^{1,2}$ and Norbert Wex$^{1}$}

\address{$^{1}\,$Max-Planck-Institut f{\"u}r Radioastronomie, 
  Auf dem H{\"u}gel 69, 53121 Bonn, Germany} 
\address{$^{2}\,$School of Physics, Peking University, Beijing 100871,
  China}
  
\eads{\mailto{lshao@pku.edu.cn}, \mailto{wex@mpifr-bonn.mpg.de}}

\begin{abstract}

Some alternative gravity theories allow the Universal matter
distribution to single out the existence of a preferred frame, which
breaks the symmetry of local Lorentz invariance (LLI) for the
gravitational interaction. In the post-Newtonian parametrization of
semi-conservative gravity theories, LLI violation is characterized by
two parameters, $\alpha_1$ and $\alpha_2$. In binary pulsars the
isotropic violation of Lorentz invariance in the gravitational sector
should lead to characteristic preferred frame effects (PFEs) in the
orbital dynamics, if the barycenter of the binary is moving relative
to the preferred frame with a velocity $\mathbf{w}$.

For small-eccentricity binaries, the effects induced by $\hat{\alpha}_1$ and  
$\hat{\alpha}_2$ (the hat indicates possible modifications by strong-field 
effects) decouple, and can therefore be tested independently. We use recent 
timing results of two compact pulsar-white dwarf binaries with known 
3-dimensional 
velocity, PSRs J1012+5307 and J1738+0333, to constrain PFEs for strongly 
self-gravitating bodies, by assuming the isotropic cosmic microwave background 
to single out a preferred frame. The time derivative of the projected semi-major 
axis is used to constrain a precession of the orbital plane around $\mathbf{w}$ 
due to PFEs. From this we derive a limit $|\hat{\alpha}_2| < 1.8 \times 10^{-4}$ 
at 95\% confidence level, which is the most constraining limit for strongly 
self-gravitating systems up to now, however still three orders of magnitude 
weaker than the best Solar system limit for the corresponding weak-field 
parameter $\alpha_2$.   

Concerning $\hat{\alpha}_1$, we propose a new, robust method to constrain this 
parameter, which avoids the probabilistic considerations inherent in previous 
methods. This method is based on the fact that a PFE-induced intrinsic 
eccentricity cannot stay unobserved during a long-term observation due to the 
significant precession of periastron in binary pulsar with short orbital 
periods. Our most  conservative result, $\hat{\alpha}_1 = -0.4^{+3.7}_{-3.1} 
\times  10^{-5}$ at 95\% confidence level from PSR J1738+0333, constitutes
a significant improvement compared to current most stringent limits obtained 
both in Solar system and binary pulsar tests.  

We also derive corresponding limits for $\hat{\alpha}_1$ and $\hat{\alpha}_2$ 
for a preferred frame that is at rest with respect to our Galaxy, and preferred 
frames that locally co-move with the rotation of our Galaxy.  

These limits will continue to improve significantly with future pulsar timing 
observations conducted at large radio telescopes.

\end{abstract}

\pacs{04.80.Cc, 11.30.Cp, 97.60.Gb}

\maketitle


\section{Introduction}

Lorentz invariance is one of the most important ingredients inherent in modern 
theoretical physics, including the standard model of particle physics and 
general relativity (GR). From a group theoretical viewpoint, it composes of two 
parts, rotational invariance and boost invariance. Rotational invariance forms a 
compact group, i.e., the ${\rm SO}(3)$ group, which can be probed throughout, 
while boost invariance forms a non-compact group, hence it, in principle, cannot 
be tested thoroughly, and deserves more scrutinies. Lorentz invariance is 
examined in particle physics to high precision \cite{kr11}, while not so well 
tested in gravitational physics \cite{wil06}, due to the challanges in 
gravitational precision experiments.

On the other hand, some alternative gravitational theories predict the existence 
of a preferred frame, which might be singled out by the matter distribution in 
our Universe or through historical relics of vectorial or tensorial vacuum 
expectation values, if gravitational interaction is mediated by a vector field 
or a second tensor field, in addition to the canonical second-rank symmetric 
tensor field \cite{wn72}. These theories include vector-metric theories 
\cite{wn72,wil93}, Einstein-\AE{}ther theories \cite{jm01}, TeVeS theories 
\cite{bek04}, and standard model extensions of gravity \cite{bk06}.

In the parametrized post-Newtonian (PPN) formalism, preferred frame effects
(PFEs) are characterized by three parameters, $\alpha_1$, $\alpha_2$ and
$\alpha_3$ \cite{wn72,wil93}. Since $\alpha_3$ also causes energy-momentum 
conservation violation, and is well constrained to very high precision (see 
e.g.~\cite{sfl+05}, where the strong-field version of $|\alpha_3|$ is 
constrained to be less than $4 \times 10^{-20}$ at 95\% confidence level, by 
using wide-orbit binary millisecond pulsars), we will not consider it further in 
this work. Experimental tests on PFEs induced by $\alpha_1$ and $\alpha_2$
are roughly divided into three catalogues, i.e., geophysical tests, Solar system 
tests, and pulsar timing tests.

Nordtvedt and Will \cite{nw72} derived possible experimental indications of a 
preferred frame for the gravitational interaction in geophysics and orbital 
motions, e.g., an anomalous 12-hour sidereal tide of the solid Earth, an 
anomalous yearly variation in the rotational frequency of the Earth, and an 
anomalous perihelion shift of the planets. By now, the best limit for $\alpha_1$ 
in the Solar system comes from Lunar Laser Ranging (LLR), that 
places a 95\% confidence limit of $\alpha_1 = (-0.7\pm 1.8)\times10^{-4}$~
\cite{mwt08}.  For the $\alpha_2$ parameter, Nordtvedt \cite{nor87} used the 
close alignment of the spin axis of the Sun and the total angular momentum 
vector of the Solar system to limit $|\alpha_2| < 2.4 \times 10^{-7}$, under the 
assumption that the above two vectors were aligned when the Solar system formed 
five billion years ago (note, $\alpha_2^{\rm Nordtvedt} = \frac{1}{2}\alpha_2$). 
Damour and Esposito-Far{\`e}se developed a method to put tight constraints onto 
the strong-field counterpart of $\alpha_1$, namely $\hat{\alpha}_1$, from timing 
experiments of small-eccentricity binary pulsars \cite{de92a}. Their calculation 
shows that the observational eccentricity vector, $\mathbf{e}(t)$, is a 
vectorial superposition of a ``rotating eccentricity'' $\mathbf{e}_R(t)$ with 
constant length $e_R$, and a fixed ``forced eccentricity'' $\mathbf{e}_F$. From 
probabilistic consideration, they were able to constrain $|\hat{\alpha}_1|$ to 
be less than $5.0 \times 10^{-4}$ (90\% C.L.). This limit has been improved by a 
factor of three in \cite{bcd96}, based on the small-eccentricity binary pulsar 
PSR~J2317+1439. Wex \cite{wex00} extended this method in statistically combining 
multiple systems, by taking care of a potential selection effect when simply 
picking the system with the most favorable parameter combination. He got a 
slightly improved result of $|\hat{\alpha}_1| < 1.2 \times 10^{-4}$ (95\% C.L.). 
In section~\ref{sec:a1}, we extend the statistically dependent method into a 
robust one which not only avoids involving probabilistic considerations 
concerning certain unobservable angles, but also gets a new constraint, 
$\hat{\alpha}_1 = -0.4^{+3.7}_{-3.1} \times 10^{-5}$ (95\% C.L.), that surpasses 
the current best constraints of both weak and strong fields.

For the $\alpha_2$ parameter, because of the tight limit of \cite{nor87}, Damour 
and Esposito-Far{\`e}se \cite{de92a} dropped the $\alpha_2$ term when calculating 
binary orbital dynamics. We stress that the limit of \cite{nor87} is obtained in 
a weak-field gravitational environment, while in the strong-field regime, like 
inside a neutron star (NS), $\alpha_2$ might take an independent value, significantly different from the Solar system value.  In fact, it was discovered that in certain classes of tensor-scalar theories of gravity, large non-perturbative strong-field deviations from GR can occur, through a phenomenon called ``spontaneous scalarization'' \cite{de93}. Although tensor-scalar theories of gravity are conservative gravity theories and do not show PFEs, it is natural to assume the possibility of similar non-perturbative effects in gravitational theories with local Lorentz invariance (LLI) violation. Hence, we feel that it is still worth to independently test, in a phenomenological approach, the strong-field counterpart of $\alpha_2$, namely $\hat{\alpha}_2$, in pulsar binary timing experiments.

Following \cite{de92a}, we calculate the $\hat{\alpha}_2$ effect for pulsar 
binaries, and find that it practically decouples from the $\hat{\alpha}_1$
effect for small orbital eccentricities ($e \ll 1$). The $\hat{\alpha}_1$ term 
tends to polarize the eccentricity vector towards a direction perpendicular to 
the orbital angular momentum and the binary barycentric velocity with respect to
the preferred frame, $\mathbf{w}$ \cite{de92a}. It causes dynamical effects 
inside the orbital plane. In contrast, the $\hat{\alpha}_2$ term imposes a 
precession of the orbital angular momentum around the direction of $\mathbf{w}$. 
It causes a change in the orbital inclination angle with respect to the line of 
sight, $i$. Consequently, $\hat{\alpha}_2$ induces a non-vanishing time
derivative of the projected semi-major axis.

Wex and Kramer \cite{wk07} developed a pulsar timing model that includes PFEs, 
by extending the widely used Damour-Deruelle timing model \cite{dd86}. Based
on this model, they analyzed the time of arrivals (TOAs) of the double pulsar, 
PSR~J0737$-$3039A/B \cite{bdp+03,lbk+04,ksm+06a,kw09}, and jointly limited 
$\hat{\alpha}_1$ and $\hat{\alpha}_2$ to be, $-0.5 < \hat{\alpha}_1 < 0.3$ and 
$-0.3 < \hat{\alpha}_2 < 0.2$, respectively.\footnote{They use $\alpha_1^*$ 
  and $\alpha_2^*$ instead of $\hat{\alpha}_1$ and  
  $\hat{\alpha}_2$ in their notation. The limit quoted above assumes the 
  preferred frame to be at rest with respect to the isotropic cosmic microwave 
  background (see the original paper for constraints on other directions).}  
Their analysis utilized two specific aspects of the double pulsar: 1) The 
measurement of the mass-ratio via the ``double-line'' nature of the system, and 
the measurement of the Shapiro delay allowed for a theory-independent 
determination 
of the effective gravitating masses of the two pulsars; 2) The large rate of 
periastron advance, $\dot{\omega} = 16.9\,{\rm deg\,yr}^{-1}$ \cite{ksm+06a}, 
would significantly change the binary orientation with respect to a preferred 
frame within just a few years, leading to distinct long-term periodic effects in 
the orbital parameters. Consequently, the double pulsar has even the potential 
to measure $\hat{\alpha}_1$ and $\hat{\alpha}_2$, if they are non-zero.  As 
emphasized in \cite{wk07}, double NS systems probe different aspects of a 
violation of LLI in the gravitational sector (interaction between two strongly 
self-gravitating bodies) from other kinds of binaries, e.g., NS-white dwarf (WD) 
binaries. Furthermore, from the simulations in \cite{wk07} one expects that by 
now the precision of the PFE test with the double pulsar has greatly improved 
compared to the numbers in \cite{wk07}.

To this point, let us briefly summarize current best limits on the LLI
violation in the weak field and strong field.
\begin{enumerate}
\item Weak field
  \begin{itemize}
  \item From LLR \cite{mwt08},
    \begin{equation}\label{llra1}
      \alpha_1 = (-0.7 \pm 1.8)\times10^{-4} \quad (95\%~{\rm C.L.})\,.
    \end{equation}
  \item From the close alignment of the spin of the Sun with the total
    angular momentum of the Solar system \cite{nor87},
    \begin{equation}\label{solara2}
      |\alpha_2|<2.4\times10^{-7}\,.
    \end{equation}
    One should be aware that this result depends on an assumption about
    the alignment of the spin of the Sun and the angular momentum of
    the Solar system right after their formation, five billion years ago. 
    LLR experiments \cite{mwt08} get a weaker limit, $\alpha_2=(1.8 \pm 5.0)
    \times 10^{-5}$ (95\% C.L.).
  \end{itemize}
\item Strong field
  \begin{itemize}
  \item From the population of small-eccentricity NS-WD binaries
    \cite{wex00},
    \begin{equation}\label{limita1}
      |\hat{\alpha}_1|<1.2\times10^{-4} \quad (95\%~{\rm C.L.})\,.
    \end{equation}
  \item From a NS-NS system, namely the double pulsar
    \cite{wk07},
    \begin{equation}\label{limita2}
      -0.3 < \hat{\alpha}_2 < 0.2 \quad (95\%~{\rm C.L.})\,.
    \end{equation}
  \end{itemize}
\end{enumerate}

In this paper, we derive the full secular dynamical evolution of a pulsar binary 
system of arbitrary eccentricity, under the influence of both $\hat{\alpha}_1$ 
and $\hat{\alpha}_2$. Afterwards, we utilize our analytical results to propose 
two new methods to constrain $ \hat{\alpha}_1$ and $\hat{\alpha}_2$, 
respectively. By using Monte Carlo simulations, we are able to get stringent 
limits from small-eccentricity NS-WD binaries, PSRs J1012+5307 
\cite{lcw+01,lwj+09} and J1738+0333 \cite{akk+12,fwe+12}, 
with measurement errors properly accounted for. 

The paper is organized as follows. In section~\ref{sec:analytical} we derive the 
orbital dynamics from a generic semi-conservative Lagrangian keeping both the $
\hat{\alpha}_1$ and the $\hat{\alpha}_2$ terms, for arbitrarily eccentric 
orbits. We find that in the limit of a small eccentricity, these two parameters 
decouple --- $\hat{\alpha}_1$ affects the evolution of the eccentricity vector 
in the orbital plane, while $\hat{\alpha}_2$ controls the precession of the 
orbital angular momentum. We introduce the isotropic cosmic microwave background 
(CMB) frame as the most important preferred frame, for our subsequent numerical
calculations in sections~\ref{sec:a2} and \ref{sec:a1}.  In 
section~\ref{sec:a2}, $|\hat{\alpha}_2|$ is derived to be less than $1.8 \times 
10^{-4}$ (95\% C.L.) from timing experiments of pulsar binaries PSRs~J1012+5307 
and J1738+0333. In section~\ref{sec:a1}, we develop a new, robust method to 
constrain $\hat{\alpha}_1$, which overcomes the need of probabilistic 
considerations inherent in the former methods. The most conservative limit, 
$\hat{\alpha}_1 = -0.4^{+3.7}_{-3.1} \times 10^{-5}$ (95\% C.L.), is derived 
from the PSR~J1738+0333 binary system. Section~\ref{sec:sum} gives the
corresponding results on $\hat{\alpha}_1$ and $\hat{\alpha}_2$ when the Galaxy 
or the local Galactic rotation are assumed to single out a preferred frame. The 
limits for these frames of reference are found to be slightly weaker than the 
ones for the CMB frame. Furthermore, we give a discussion on the strong-field 
aspects of our tests, discuss future improvements of these two tests, and 
briefly summarize the results of the paper.


\section{Binary dynamics of the semi-conservative Lagrangian}
\label{sec:analytical}

We consider the two-body dynamics of a binary system consisting of a
pulsar with mass $m_p$ and its companion with mass $m_c$. In the
presence of a preferred reference frame, the orbital motion of such a
system is described by a two-body non-boost-invariant Lagrangian
\cite{wil93,de92a}
\begin{equation}\label{fullL}
  L = L_{\hat{\beta},\hat{\gamma}} + L_{\hat{\alpha}_1} +
  L_{\hat{\alpha}_2} \,.
\end{equation}
The Lagrangian~(\ref{fullL}) consists of
$L_{\hat{\beta},\hat{\gamma}}$, the post-Newtonian (PN) terms from
GR and its minimal extensions characterized by the (strong-field)
Eddington-Robertson-Schiff parameters, $\hat{\beta}$ and
$\hat{\gamma}$,
\begin{eqnarray}
\hspace{-2cm}L_{\hat{\beta},\hat{\gamma}} &=& 
    -m_pc^2 \sqrt{1-\frac{(\mathbf{v}_p^0)^2}{c^2}}
    -m_cc^2 \sqrt{1-\frac{(\mathbf{v}_c^0)^2}{c^2}}
    +\frac{\hat{G}m_pm_c}{r} \left[1 + 
  \frac{(\mathbf{v}_p^0)^2 + (\mathbf{v}_c^0)^2}{2c^2}  
   \right.
 \nonumber\\
\hspace{-2cm} && \left. - \frac{3(\mathbf{v}_p^0 \cdot
  \mathbf{v}_c^0)}{2c^2} 
  -\frac{({\mathbf{n}}\cdot\mathbf{v}_p^0)
  ({\mathbf{n}}\cdot\mathbf{v}_c^0)}{2c^2}
   + \hat{\gamma}\,
  \frac{(\mathbf{v}_p^0 - \mathbf{v}_c^0)^2 }{c^2}
  - (2\hat{\beta}-1) \frac{\hat{G}M}{2c^2r} \right] \,,
\end{eqnarray}
and the velocity-dependent, non-boost-invariant terms,
related to non-vanishing ${\hat{\alpha}_1}$ and ${\hat{\alpha}_2}$,
\begin{eqnarray}
  L_{\hat{\alpha}_1} &=& 
    -\hat{\alpha}_1 \, \frac{\hat{G}m_pm_c}{r} \,
     \frac{(\mathbf{v}_p^0 \cdot \mathbf{v}_c^0)}{2c^2} \,,\\
  L_{\hat{\alpha}_2} &=& 
    \hspace{0.8em} \hat{\alpha}_2 \, \frac{\hat{G}m_pm_c}{r} \,
        \frac{(\mathbf{v}_p^0 \cdot \mathbf{v}_c^0) -
        ({\mathbf{n}}\cdot\mathbf{v}_p^0)
        ({\mathbf{n}}\cdot\mathbf{v}_c^0) }{2c^2} \,,
\end{eqnarray}
where $M \equiv m_p + m_c$, $r \equiv |\mathbf{r}|$ is the coordinate separation 
of two components, ${\mathbf{n}}\equiv \mathbf{r}/r$, $\mathbf{v}^0$ denotes the 
``absolute'' velocity with respect to the preferred frame, and $c$ is the speed 
of light.  In the above Lagrangian, we add a ``hat'' onto the notations of 
$\gamma$, $\beta$, $\alpha_1$, $\alpha_2$, and also the gravitational constant 
$G$, to underline the fact that we are dealing with the PN parameters associated 
with compact objects, where strong-field effects might contribute to these 
values, making them different from their counterparts in the weak field. The 
specific dependence on the strong-field contributions depends on the 
gravitational theories under consideration. In GR, one finds $\hat{G} = G$, 
$\hat{\beta} =\hat{\gamma} = 1$, and $\hat{\alpha}_1 = \hat{\alpha}_2 = 0$.

The Lagrangian~(\ref{fullL}) can be obtained from a more generalized,
semi-conservative Einstein-Infeld-Hoffmann Lagrangian\footnote{A
  ``semi-conservative'' Lagrangian corresponds to a gravity theory
  that possesses conservation laws for the total energy and momentum.
  Any theory that is based on an invariant action principle is
  ``semi-conservative'' \cite{wil06}.}  in the calculations of
\cite{wil93} and \cite{wk07}, by setting
$\mathcal{A}_p = \mathcal{A}_c = 1$, $\mathcal{G} = \hat{G}/G$,
$\mathcal{B}/\mathcal{G} = \frac{1}{3} (2\hat{\gamma}+1)$,
$\mathcal{C}/\mathcal{G} = \frac{1}{7} (4\hat{\gamma}+
\hat{\alpha}_1 -\hat{\alpha}_2+3)$, $\mathcal{E}/\mathcal{G} =
\hat{\alpha}_2 + 1$, $X_c\mathcal{D}_p/\mathcal{G}^2 +
X_p\mathcal{D}_c/\mathcal{G}^2 = 2\hat{\beta} - 1$, where $X_p\equiv
m_p/M$ and $X_c \equiv m_c/M$. 

The assumption $\mathcal{A}_p = \mathcal{A}_c = 1$, made in this paper, 
requires additional justification.  These parameters are equal to one 
at first post-Newtonian order \cite{wil93}, but for strongly self-gravitating 
bodies, e.g.~NSs, they could significantly deviate from one (e.g.\ in 
Einstein-\AE{}ther theory \cite{fos07}). The subject of this paper are pulsars 
with low-mass WD companions, and therefore $\mathcal{A}_c \simeq 
1$. Concerning $\mathcal{A}_p$, it enters the secular changes of the relevant orbital parameters only as addition to the parameters $\hat{\alpha}_1$ and 
$\hat{\alpha}_2$ (see \cite{wk07} for details). In the 
first case the $\mathcal{A}_p$ contribution is multiplied by a factor of 
$2X_c^2$ and in the second case by a factor $X_c$. Since $X_c \sim 0.1$, for our 
binary systems, in both cases the $\mathcal{A}_p$ contributions are expected to 
be small with respect to the $\hat{\alpha}_1$ and $\hat{\alpha}_2$ terms.

Besides the PPN parameters $\beta$, $\gamma$, $\alpha_1$ and $\alpha_2$, 
semi-conservative theories of gravity could have a non-zero Whitehead term, 
characterized by $\xi$ \cite{wil93}. It reflects preferred-location effects, 
such as an anisotropy in the local gravitational constant caused by an external 
gravitational potential. Various well-motivated (fully conservative and semi-
conservative) gravity theories have $\xi = 0$ (for instance, see Table 3 of 
\cite{wil06}). Therefore, we will only include $\hat{\beta}$, $\hat{\gamma}$, 
$\hat{\alpha}_1$, and $\hat{\alpha}_2$ in our following 
discussion, and ignore a potential strong-field counterpart of the Whitehead 
term. We note in passing, that for small-eccentricity binaries, the presence of 
a Whitehead term only changes the $\hat\alpha_2$ test, which then 
constrains a combination of $\hat\alpha_2$ and $\hat\xi$. The $\hat\alpha_1$
test, on the other hand, remains unchanged. This can be readily seen from 
(8.73) in \cite{wil93}.


\subsection{Orbital dynamics in the presence of PFEs}

Using the Euler-Lagrange equations and the post-Galilean transformations 
\cite{cc67}, we can get the relative acceleration for a pulsar binary system in 
the comoving frame, whose center of mass moves relative to the preferred frame 
with a velocity $\mathbf{w}$,
\begin{equation}
  \ddot{\mathbf{r}} \equiv \ddot{\mathbf{r}}_p - \ddot{\mathbf{r}}_c 
             = \mathbf{A}^{\rm (N)} + \mathbf{A}^{\rm (PN)}/c^2 +
  \mathbf{A}^{(\mathbf{w})}/c^2 \,,
\end{equation}
where $\mathbf{A}^{\rm (N)}$ is the ``Newtonian'' acceleration,
$\mathbf{A}^{\rm (N)} = -\hat{G}M {\mathbf{n}} /r^2$,
$\mathbf{A}^{\rm (PN)}/c^2$ is the first PN acceleration
without $\mathbf{w}$-dependent contributions, and
$\mathbf{A}^{(\mathbf{w})}/c^2$ is the additional acceleration from
the motion of the binary system with respect to the preferred
frame. For expressions of these accelerations, see \cite{wil93,de92a,wk07}.


\begin{figure}
\begin{center}
\includegraphics[width=10cm]{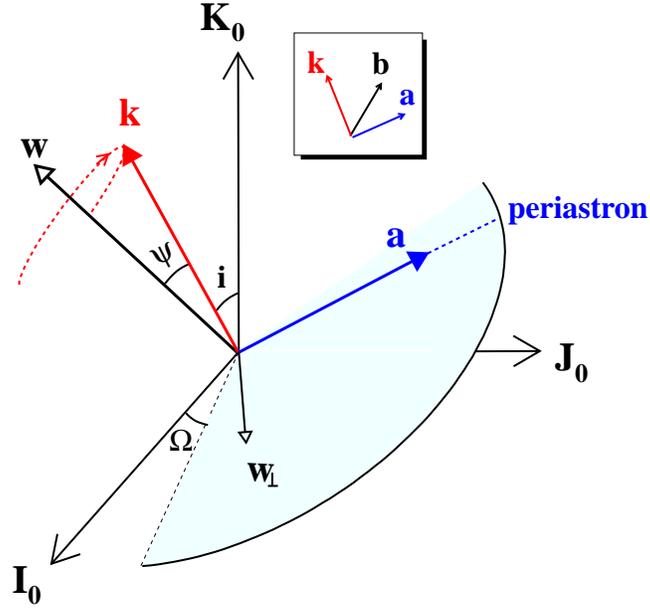}
\end{center}
\caption{\label{geometry}
  Illustration of the geometry of the binary system, and the notation used in 
  the paper. The coordinate system $({\mathbf{I}}_0, {\mathbf{J}}_0, 
  {\mathbf{K}}_0)$ corresponds to $(\vec{I}_0, \vec{J}_0, \vec{K}_0)$ in 
  figure~1 or $(\mathbf{I}_0, \mathbf{J}_0, \mathbf{K}_0)$ in the text of
  \cite{dt92}, and the coordinate system $({\mathbf{a}}, {\mathbf{b}},
  {\mathbf{k}})$ corresponds to $(\mathbf{a}, \mathbf{b}, \mathbf{c})$ in 
  \cite{de92a}. $\mathbf{w}$ is the velocity of the binary system with respect 
  to the preferred frame, while $\mathbf{w}_\perp$ is its projection into the 
  orbital plane.}
\end{figure}


As is well known, the acceleration $\mathbf{A}^{\rm (PN)}/c^2$
produces a secular advance of the longitude of periastron,
\begin{equation}\label{omegaR}
    \dot{\omega}_{\rm PN} = 
      \frac{3 {\cal V}_O^2 {\mathcal{F}}}{c^2(1-e^2)}\,n_b \,,
\end{equation}
where
\begin{eqnarray}
  {\cal V}_O &\equiv& (\hat{G}Mn_b)^{1/3} \,,\\
  {\mathcal{F}} &\equiv& 
    \frac{1}{3}(2+2\hat{\gamma}-\hat{\beta}) +
    \frac{1}{6}(2\hat{\alpha}_1 - \hat{\alpha}_2)X_pX_c \,,
\end{eqnarray}
and $n_b = 2\pi/P_b$ is the orbital frequency of the binary system.\footnote{
  The velocity ${\cal V}_O$, which corresponds to $\beta_O c$ in \cite{dt92}, 
  should not be confused with $v_0 \equiv \mathcal{V}_O/\sqrt{1-e^2}$ in 
  (17--21) of \cite{wk07}.}  
The ``characteristic'' velocities in GR, $\mathcal{V}_O^{\rm (GR)}$, for
pulsar binaries PSRs~J1012+5307 and J1738+0333, are given in
table~\ref{tab:pars}. In GR, $\mathcal{F}=1$.  Because $\mathbf{A}^{\rm (PN)} / 
c^2$ lies in the orbital plane, it has no effect on the
longitude of the ascending node $\Omega$, and the orbital inclination
angle $i$ (see figure~\ref{geometry} for illustration of these angles). In 
addition, $\mathbf{A}^{\rm (PN)}/c^2$ is verified to have no effect on $e$, the 
length of the eccentricity vector, and $a$, the semi-major axis of the relative
orbit (cf.\ (\ref{a1dedt}) below, with ${\bf w}=0$)

As for the acceleration from PFEs, $\mathbf{A}^{(\mathbf{w})}/c^2$, Damour and 
Esposito-Far{\`e}se \cite{de92a} worked out the influence of 
$\hat{\alpha}_1$-related terms on the orbital evolution. After averaging over 
one orbital period, they found for changes in the semi-major axis $a$, 
$\mathbf{l} \equiv \sqrt{1-e^2} \, {\mathbf{k}}$, and the eccentricity vector $
\mathbf{e} \equiv e \, {\mathbf{a}}$,
\begin{eqnarray}
  \fl\left\langle \frac{\textrm{d}a}{\textrm{d}t}
    \right\rangle_{\mathrm{PN}+\hat{\alpha}_1} &=& 0 \,, 
    \label{a1dadt}\\
  \fl\left\langle \frac{\textrm{d}\mathbf{l}}{\textrm{d}t}
    \right\rangle_{\mathrm{PN}+\hat{\alpha}_1} &=& 
    \frac{\hat{\alpha}_1}{2c^2} \, \frac{q-1}{q+1}\,n_b\mathcal{V}_O \, 
      e F_e  \, ({\mathbf{b}} \times \mathbf{w}) \,,
    \label{a1dldt}\\
  \fl\left\langle \frac{\textrm{d}\mathbf{e}}{\textrm{d}t}
    \right\rangle_{\mathrm{PN}+\hat{\alpha}_1} &=& 
    e \, \dot{\omega}_{\rm PN} \, {\mathbf{b}} +
    \frac{\hat{\alpha}_1}{2c^2} 
    \, \frac{q-1}{q+1}\,n_b\mathcal{V}_O F_e 
    \left( \sqrt{1-e^2} \, w_a \, {\mathbf{a}} + w_b \, {\mathbf{b}} - 
     \frac{e^2 w_k}{\sqrt{1 - e^2}} \, {\mathbf{k}}\right) \,,
    \label{a1dedt} 
\end{eqnarray}
where $q\equiv m_p/m_c$ is an observable quantity, due to the additional optical 
information for the two pulsar binaries used in this paper. The three unit 
vectors (${\mathbf{a}}$, ${\mathbf{b}}$, ${\mathbf{k}}$) form a right-handed 
triad of a coordinate system with its origin at the center of mass of the binary 
system, where ${\mathbf{a}}$ points to the position of periastron, 
${\mathbf{k}}$ points along the orbital angular momentum, and ${\mathbf{b}} 
\equiv {\mathbf{k}} \times {\mathbf{a}}$, as illustrated in 
figure~\ref{geometry}. Furthermore,
\begin{equation}
  F_e \equiv \frac{1}{1+\sqrt{1-e^2}} \,,
\end{equation} 
is a function that, for bound orbits ($0 \le e < 1$), takes a value
in the interval $\left[\frac{1}{2},1\right)$.

In addition to the contribution from the $ \hat{\alpha}_1$ and PN
terms, we find that for the change caused by the $ \hat{\alpha}_2$
terms, after averaging over one orbital period,
\begin{eqnarray}
  \fl\left\langle \frac{\textrm{d}a}{\textrm{d}t}
    \right\rangle_{\hat{\alpha}_2} &=& 0 \,, 
    \label{a2dadt} \\
  \fl\left\langle \frac{\textrm{d}\mathbf{l}}{\textrm{d}t}
    \right\rangle_{\hat{\alpha}_2} &=& 
    \frac{\hat{\alpha}_2}{c^2} n_b F_e \left(
    w_k {\mathbf{k}} +  e^2F_e \, w_b 
    {\mathbf{b}}  \right) \times \mathbf{w} \,,
    \label{a2dldt} \\
  \fl\left\langle \frac{\textrm{d}\mathbf{e}}{\textrm{d}t}
    \right\rangle_{\hat{\alpha}_2} &=& 
    \frac{\hat{\alpha}_2}{c^2} n_b F_e
    \left( F_e\sqrt{1 - e^2}\, w_a w_b \, {\mathbf{a}} 
     -F_e\,\frac{w_a^2 - w_b^2}{2} \, {\mathbf{b}}  
     +w_b w_k \, {\mathbf{k}} \right) e \,, 
     \label{a2dedt}
\end{eqnarray}
where $(w_a,w_b,w_k)$ are the coordinate components of $\mathbf{w}$ in the 
$({\mathbf{a}},{\mathbf{b}},{\mathbf{k}})$ system.

From (\ref{a1dadt}) and (\ref{a2dadt}) we can see that, as expected, to first 
order there is no change in the semi-major axis of the orbit from PFEs.


\subsection{Small-eccentricity orbits and PFEs}

The coupled differential equations above simplify considerably for small 
eccentricities. When $e \ll 1$, one finds $F_e \simeq 1/2$ and $\mathbf{l} 
\simeq {\mathbf{k}}$. To leading order in the (numerically) relevant 
contributions, (\ref{a1dldt}) and (\ref{a2dldt}) become
\begin{eqnarray}
  \fl\left\langle\frac{\textrm{d}{\mathbf{k}}}{\textrm{d}t}
     \right\rangle_{\mathrm{PN}+\hat{\alpha}_1} &\simeq& 0 \,,
  \label{ell1a1dldt} \\
 \fl\left\langle \frac{\textrm{d}{\mathbf{k}}}{\textrm{d}t}
    \right\rangle_{\hat{\alpha}_2} &\simeq& \frac{\hat{\alpha}_2}{2c^2} n_b
    w_k {\mathbf{k}} \times \mathbf{w} \,,
 \label{ell1a2dldt} 
\end{eqnarray}
and (\ref{a1dedt}) and (\ref{a2dedt}) simplify to 
\begin{eqnarray}
  \fl\left\langle \frac{\textrm{d}\mathbf{e}}{\textrm{d}t}
     \right\rangle_{\mathrm{PN}+\hat{\alpha}_1} &\simeq& 
     e \, \dot{\omega}_{\rm PN} \, {\mathbf{b}} +
     \frac{\hat{\alpha}_1}{4c^2} \, \frac{q-1}{q+1} \,
     n_b {\cal V}_O \, \mathbf{w}_\perp \,,
 \label{ell1a1dedt} \\
 \fl\left\langle \frac{\textrm{d}\mathbf{e}}{\textrm{d}t}
\right\rangle_{\hat{\alpha}_2} &\simeq& 0 \,,
 \label{ell1a2dedt}
\end{eqnarray}
where $w \equiv|\mathbf{w}|$ and $\mathbf{w}_\perp \equiv w_a
{\mathbf{a}} + w_b {\mathbf{b}}$ is the projection of
$\mathbf{w}$ into the orbital plane. Above four equations have been derived, 
under the consideration that $e \lesssim 10^{-6}$, $w^2/c^2 \sim 
{\cal V}_O w/c^2 \sim {\cal V}_O^2/c^2 \sim 10^{-6}$, for the NS-WD systems 
which are to be used in our calculations, i.e., PSRs J1012+5307 and J1738+0333 (see table~\ref{tab:pars}). 

From (\ref{ell1a1dldt}) and (\ref{ell1a1dedt}), Damour and
Esposito-Far{\`e}se \cite{de92a} worked out the eccentricity vector
evolution under the influence of the PN and $\hat{\alpha}_1$ terms,
which turns out to be a superposition of a PN-induced precessing eccentricity
$\mathbf{e}_R(t)$, and a constant ``forced eccentricity''
$\mathbf{e}_F$, introduced by $\hat{\alpha}_1$. 

In terms of a geometrical interpretation of the time evolution of the
orbital eccentricity, the physical consequence of $\hat{\alpha}_1$
was extensively studied \cite{de92a,bcd96,wex00}. The physical
consequence of $\hat{\alpha}_2$ for small-eccentricity binary
systems is readily derived from (\ref{ell1a2dldt}), which shows that a
non-zero $\hat{\alpha}_2$ causes a precession of the orbital angular
momentum around the fixed direction $\mathbf{w}$ with an angular
frequency,
\begin{equation}\label{omegaprec}
  \Omega^{\mathrm{prec}}_{\hat{\alpha}_2} = -\frac{\hat{\alpha}_2}{2} n_b
  \left( \frac{w}{c}\right)^2 \cos \psi \,,
\end{equation}
where $\psi$ is the angle between ${\mathbf{k}}$ and $\mathbf{w}$
(see figure~\ref{geometry} for an illustration of the orbital geometry and the 
orbital angular momentum precession). To leading order, this precession is 
purely determined by $\hat{\alpha}_2$ (see (\ref{ell1a1dldt})).

The precession (\ref{omegaprec}) induces a secular change of the
projected semi-major axis of the pulsar orbit. The rate of change is
given by
\begin{equation}\label{a2xdot}
  \left(\frac{\dot{x}}{x}\right)_{\hat{\alpha}_2} = 
  -\frac{\hat{\alpha}_2}{4} n_b \left(\frac{w}{c}\right)^2 
            \cot i \, \sin2\psi \, \cos \vartheta \,,
\end{equation}
where $\vartheta$ is the angle between $\mathbf{w}_\perp$ and the direction of 
ascending node. In section~\ref{sec:a2}, we will apply (\ref{a2xdot}) to 
constrain ${\hat{\alpha}_2}$ from two relativistic small-eccentricity NS-WD 
binaries, namely, PSRs~J1012+5307 and J1738+0333.


\subsection{The preferred frame}
\label{sec:pf}

As the most natural preferred frame for our following calculations, we choose 
the frame determined by the isotropic CMB, like this is generally done in the 
literature on preferred-frame tests. To use other frames, the generalization is 
straightforward. As an example, in section~\ref{sec:sum} we also present limits 
on ${\hat{\alpha}_1}$ and ${\hat{\alpha}_2}$ for which the Galaxy or the
local Galactic rotation is assumed to determine the preferred frame.

From the five-year Wilkinson Microwave Anisotropy Probe (WMAP) operations, a CMB 
dipole measurement of $3.355 \pm 0.008~\textrm{mK}$ was obtained, which implies 
a peculiar velocity of the Solar system barycenter (SSB) with respect to the CMB frame of $|\mathbf{v}_{\mathrm{SSB-CMB}}| = 369.0 \pm 0.9~\textrm{km\,s}^{-1}$,
in the direction of Galactic longitude and latitude $(l,b) = (263.99^\circ
\pm 0.14^\circ, 48.26^\circ \pm 0.03^\circ)$ \cite{hwh+09}. Results from 
the seven-year WMAP observations are unchanged \cite{jbd+11}. The binary velocity with respect to the preferred frame is $\mathbf{w} = 
\mathbf{v}_{\mathrm{PSR-SSB}} + \mathbf{v}_{\mathrm{SSB-CMB}}$, where
$\mathbf{v}_{\mathrm{PSR-SSB}}$ is the 3-dimensional (3D) 
motion of the pulsar binary system, with respect to the SSB. For PSRs J1012+5307 
and J1738+0333, $\mathbf{v}_{\mathrm{SSB-CMB}}$ can be derived from a 
combination of the distance and proper motion measurements from radio timing, 
and the radial velocity obtained from spectral observations of the WD.


\section{New constraints on $\hat{\alpha}_2$}
\label{sec:a2}

To constrain $\hat\alpha_2$ from binary pulsar observations, we start from
(\ref{a2xdot}), where the orbital frequency, $n_b = 2\pi/P_b$, and the projected 
semi-major axis, $x$, are observable Keplerian parameters, while the time 
derivative of $x$, $\dot{x}$, belongs to the set of phenomenological 
post-Keplerian parameters \cite{dd86,dt92}. They are obtained with high 
precision from radio timing 
observations. In (\ref{a2xdot}), we also need the inclination of the binary 
orbit with respect to the line of sight $i$.  For the binaries of this paper, 
$i$ can be determined (modulo the ambiguity of $i \rightarrow 180^\circ - i$) 
from the mass function, leading to
\begin{equation}\label{eq:sini}
  \sin i = \frac{c x n_b}{\mathcal{V}_O}\,(q+1) \,.
\end{equation}
The companion mass $m_c$ is inferred from spectroscopic and photometric studies 
of the WD companion using well tested atmospheric model for such WDs 
\cite{cgk98,akk+12}. The pulsar mass $m_p$ is determined from the mass ratio of 
the pulsar and its companion, $q \equiv m_p / m_c$, which is inferred from the 
radial velocity and the orbital parameters of the binary system 
\cite{cgk98,akk+12}. Unfortunately, the information is not sufficient to 
calculate $\sin i$, since $\mathcal{V}_O$ contains, besides the known total mass
$M$, the effective gravitational constant $\hat{G}$ which is {\it a priori} 
unknown if one does not specify a given gravity theory. In principle, 
strong-field modification could lead to a significant deviation of $\hat{G}$ 
from the GR value, $G$. Such modifications, on the other hand, are expected to 
be accompanied by a significant amount of dipolar gravitational radiation (as an 
example, see \cite{de92a} for the case of tensor-scalar theories of gravity), 
which is neither the case in PSR~J1012+5307 \cite{lwj+09} nor in
PSR~J1738+0333 \cite{fwe+12}. Consequently, for the required precision in $i$, 
we can safely assume $\hat{G} \simeq G$ in (\ref{eq:sini}).

In order to fully determine the orientation of the binary with respect to 
$\mathbf{w}$ ($\psi$ and $\vartheta$ in (\ref{a2xdot})), one also needs the 
longitude of the ascending node, $\Omega$, an angle which (in most cases) is not
measurable from pulsar timing experiments. Consequently, in our $\hat\alpha_2$ 
tests we will treat $\Omega$ as a random variable uniformly distributed between 
$0^\circ$ and $360^\circ$. This, however, will require probabilistic arguments 
in order to exclude those (small) ranges of $\Omega$ where $\hat\alpha_2$ would 
practically be unconstrained.


\subsection{PSR~J1012+5307}


\begin{figure}
\begin{center}
\includegraphics[width=14cm]{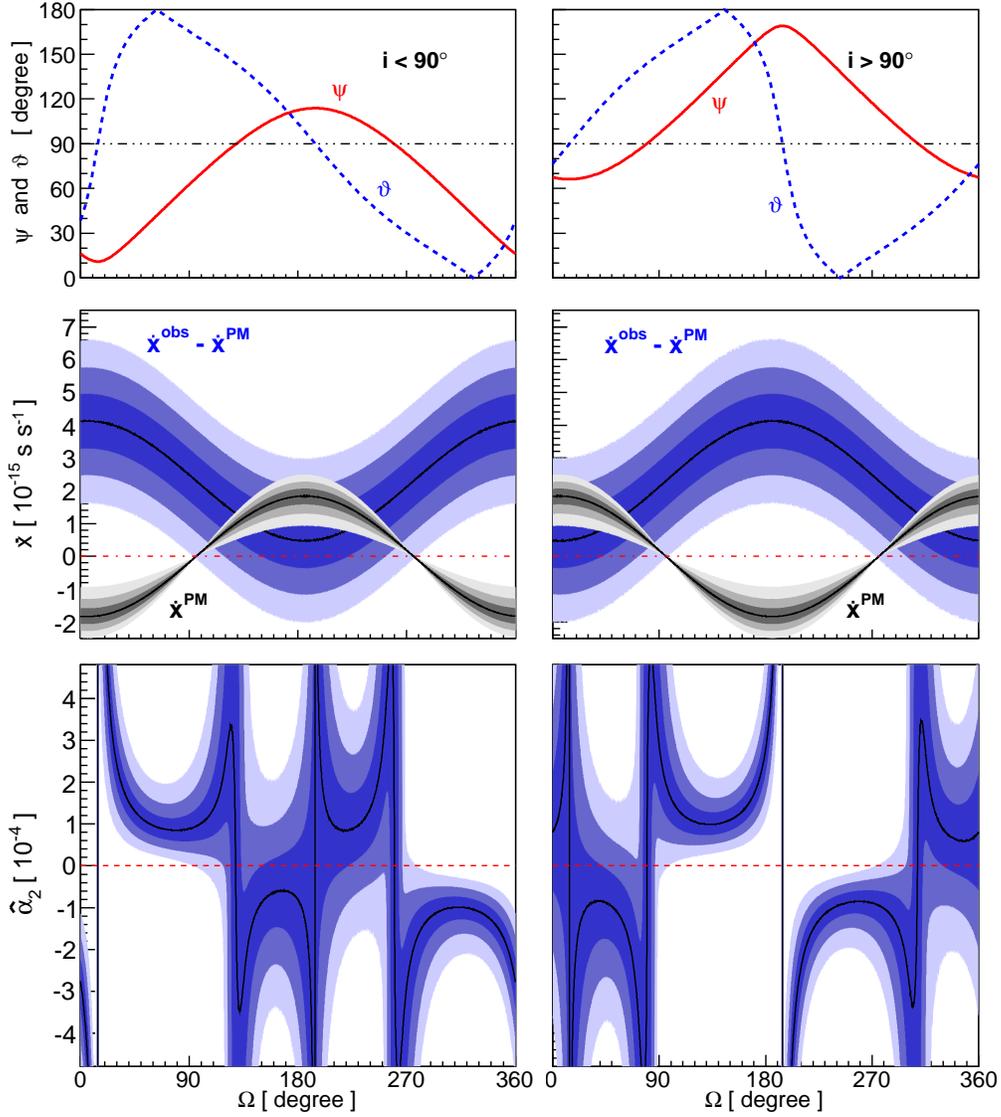}
\end{center}
\caption{\label{fig:a2_J1012}
{\it Upper:} 
  Illustration of two important angles of the PSR~J1012+5307 binary in our   
  calculation, as a function of the unobservable longitude
  of the ascending node $\Omega$: $\psi$, the angle between
  $\mathbf{w}$ and the orbital angular momentum, and $\vartheta$,
  the angle between $\mathbf{w}_\perp$ and the direction of ascending
  node. 
{\it Middle:} Proper motion contribution to $\dot{x}$, namely
  $\dot{x}^{\rm PM}$ (gray), and the residual $\dot{x}$, namely
  $\dot{x}^{\rm obs}-\dot{x}^{\rm PM}$ (blue).  
{\it Lower:} Derived $\hat{\alpha}_2$ from the residual $\dot{x}$. 
  Different contours correspond to $68\%$, $95\%$, and $99.7\%$ 
  confidence levels. The left column is for the $i<90^\circ$ branch, and the 
  right column is for the $i>90^\circ$ branch.}
\end{figure}


PSR~J1012+5307 is a small-eccentricity NS-WD binary system, with an orbital 
period of $\sim 14.5$\,h. The pulsar was discovered in 1993 with the 76-m Lovell 
radio telescope at
Jodrell Bank~\cite{nll+95}, and optical observations revealed its
companion being a helium WD \cite{llfn95}. Callanan~\etal~\cite{cgk98}
measured a systemic radial velocity of $44\pm8\,\textrm{km\,s}^{-1}$
relative to the SSB, the mass ratio $q = 10.5 \pm 0.5$, and the 
companion mass $m_c = 0.16 \pm 0.02\,\textrm{M}_\odot$.

Lange \etal~\cite{lcw+01} used 4-year Effelsberg 100-m radio telescope
timing data and 7-year 76-m Lovell telescope timing data to derive a
set of pulsar timing parameters, by utilizing the low eccentricity
binary timing model \texttt{ELL1}. These timing results were used to put
stringent limits on the emission of dipolar gravitational radiation by this 
NS-WD system. Most recently, Lazaridis \etal~\cite{lwj+09} updated the timing 
parameters by using 15 years of observations from the European Pulsar Timing 
Array (EPTA) network, consisting of the four radio-telescopes Effelsberg 
(Germany), Jodrell Bank (UK), Westerbork (the Netherlands) and Nan{\c c}ay 
(France). 

The short orbital period and the measurement of its spatial systemic velocity 
make the PSR~J1012+5307 system particularly interesting for studies of PFEs
and tests of the corresponding parameters.
To utilize (\ref{a2xdot}), the full information of $\mathbf{w}$ and
the orientation of the orbital plane is needed. To get $\mathbf{w}$,
we calculate the binary velocity with respect to SSB from
the transverse velocity obtained from radio timing and the
radial velocity measurement via spectroscopy of the WD companion. With these
measurements at hand one can compute $\mathbf{w} = \mathbf{v}_{\mathrm{PSR-SSB}} 
+ \mathbf{v}_{\mathrm{SSB-CMB}}$. The inclination of the orbital plane,
$i$, is calculated from (\ref{eq:sini}),
with a sign ambiguity between $i<90^\circ$ and $i>90^\circ$. Hence we
have two branches of solution. Moreover, pulsar timing experiments
generally give no information on the longitude of the ascending node
$\Omega$. We sample it in the range $[0,360^\circ)$. Figure~\ref{fig:a2_J1012}
illustrates $\psi$ (angle between $\mathbf{w}$ and the orbital angular momentum) 
and $\vartheta$ (angle between the ascending node and $\mathbf{w}_\perp$), as a 
function of $\Omega$. In the figures of $\psi$ and $\vartheta$,
measurement uncertainties of $m_c$, $q$, and the proper motion are not
included. However, our simulations to constrain $\hat{\alpha}_2$,
which are to be discussed below, take full account of all measurement 
uncertainties.

To look into the change of $x$ induced by the $\hat{\alpha}_2$ term,
we should separate other potential effects from the measured
$\dot{x}$.  A change of $x$ can come from various astrophysical and
gravitational effects \cite{dt92,lk05}.  All effects that cause a
change to the semi-major axis of the system, like gravitational wave
damping and component mass loss, can be constrained observationally
via the observed $\dot{P}_b^{\mathrm{obs}}$ (see table~\ref{tab:pars}). 
In fact, we can re-write (8.76) in \cite{lk05} to
\begin{eqnarray}\label{effectsonxdot}
  \hspace{-2cm}\left(\frac{\dot{x}}{x}\right)^{\mathrm{obs}} &=&
  \frac{2}{3} \left(\frac{\dot{P}_b}{P_b}\right)^{\mathrm{obs}}
  - \frac{\dot{D}}{3D}
  + \left(\frac{\dot{x}}{x}\right)^{\mathrm{PM}}  
  + \frac{\textrm{d} \varepsilon_A}{\textrm{d} t} 
  + \left(\frac{\dot{x}}{x}\right)^{\mathrm{SO}}
  + \left(\frac{\dot{x}}{x}\right)^{\mathrm{planet}} \,.
\end{eqnarray}
The remaining terms are due to a change in the Doppler factor $D$, the proper 
motion of the binary system, a change in the aberration due to a change in the 
pulsar-spin orientation, a change in the orbital inclination due to spin-orbit 
coupling effects, and finally a mass distribution in the vicinity of the system.  
Before calculating $\hat{\alpha}_2$ from (\ref{a2xdot}), these influences on 
$\dot{x}$ should be subtracted. We will discuss them term by term in the 
following.

Using the measured quantities of PSR~J1012+5307 in table~\ref{tab:pars}, the 
first term on the right-hand side of (\ref{effectsonxdot}) can be estimated to 
be $\dot{x}^{\dot{P}_b}\sim 4 \times 10^{-19}\,{\rm s\,s}^{-1}$, which is four 
orders of magnitude smaller than the
relevant scale.

The second term, $-\dot{D}/3D$, includes contributions from the Galactic
acceleration of the binary system, and the Shklovskii effect that is induced by 
the transverse proper motion \cite{shk70}. One finds 
\cite{dt91}
\begin{equation}\label{ddot}
  -\frac{\dot{D}}{D} = 
  \frac{1}{c} {\mathbf{K}}_0 \cdot
  (\mathbf{g}_{\mathrm{PSR}} - \mathbf{g}_{\mathrm{SSB}}) + 
  \frac{v_T^2}{cd} \,,
\end{equation}
where ${\mathbf{K}}_0$ is the unit vector pointing from the SSB to the binary 
pulsar, as defined before, $d$ is the pulsar distance from the SSB, $v_T =
d\sqrt{\mu_\alpha^2+\mu_\delta^2}$ is the transverse velocity of the system with 
respect to the SSB, $\mathbf{g}_{\mathrm{PSR}}$ and $\mathbf{g}_{\mathrm{SSB}}$ 
are the Galactic accelerations of the binary and the SSB, respectively. The 
contributions from the Galactic acceleration and the Shklovskii effect to 
$\dot{x}$ are of order $2\times10^{-20}\,{\rm s\,s}^{-1}$ and 
$3 \times 10^{-19} \, {\rm s\,s}^{-1}$, respectively. Hence, both contributions 
are negligible \cite{lwj+09}.

The third term of (\ref{effectsonxdot}) is a variation of $x$ caused
by a change of the orbital inclination $i$, due to the proper motion
of the binary system \cite{ajrt96,kop96},
\begin{equation}\label{xdotpm}
  \left(\frac{\dot{x}}{x}\right)^{\rm PM} 
    = ( -\mu_\alpha \sin\Omega + \mu_\delta \cos\Omega ) \cot i \,.
\end{equation}
This contribution is not negligible. The contribution from the proper
motion effect is depicted in the middle panels of
figure~\ref{fig:a2_J1012} for $i < 90^\circ$ (left) and $i > 90^\circ$
(right), respectively, as a function of the unknown longitude of the
ascending node $\Omega$. In \cite{lwj+09} the $\dot{x}$ measurement was used to 
constrain $\Omega$, by
assuming that the measured $\dot{x}$ is solely caused by the proper
motion effect~(\ref{xdotpm}). They got constraints on $\Omega$, by
requiring $\dot{x}^{\rm obs} = \dot{x}^{\rm PM}$ (see (11--14) in
\cite{lwj+09}).  In our test we have to keep full ignorance of $\Omega$, 
and cannot assume $\dot{x}^{\rm obs} = \dot{x}^{\rm PM}$. The residual 
$\dot{x}$, after subtracting of the
contribution from binary proper motion, is also plotted in the same
figure.  The intersections of $\dot{x}^{\rm obs} - \dot{x}^{\rm PM}$
and the horizontal null lines in the figure correspond to the limit on
$\Omega$ obtained in \cite{lwj+09}.

The fourth term in (\ref{effectsonxdot}) is due to the varying
aberration caused by geodetic precession of the pulsar spin axis
\cite{dt92}. For a nearly circular orbit one finds
\begin{equation}
  \frac{\textrm{d} \varepsilon_A}{\textrm{d} t} \simeq -\frac{P}{P_b} \,
  \frac{\cot \lambda \sin 2\eta + \cot i \cos \eta}{\sin \lambda} \, 
  \Omega_{\rm geod} \,,
\end{equation}
where $P$ is the pulsar spin period, $\lambda$ and $\eta$ are
positional angles of the pulsar spin vector (see figure~1 in
\cite{dt92} for details).  In GR, the geodetic precession rate for a
nearly circular orbit is given by \cite{bo75}
\begin{equation}
  \Omega_{\rm geod} \simeq
    \frac{3 + 4 q}{2(1 + q)^2} \,
    \left(\frac{\mathcal{V}_O^{\rm (GR)}}{c}\right)^2 \, n_b \,.
\end{equation}
Consequently, from the timing parameters in table~\ref{tab:pars} one finds that
the ${\textrm{d} \varepsilon_A}/{\textrm{d} t}$ term produces a change of
$x$ of order $10^{-18}\,{\rm s\,s}^{-1}$ for typical spin orientations. Hence, 
it is negligible in our case, unless there is a deviation from GR by at least a 
factor of 100, which we consider as highly unlikely, as such a large deviation 
of gravity in this system is clearly not seen in the gravitational wave emission 
\cite{lwj+09}. Moreover, PSR~J1012+5307 is a highly recycled pulsar, and 
therefore its spin axis is expected to be nearly aligned with the orbital 
angular momentum (i.e.~$\eta \simeq -90^\circ$ and $\lambda \simeq i$), which 
greatly suppresses this effect anyway.

The classical spin-orbital coupling, due to the quadrupole moment of the 
companion star, can change the inclination of the orbital plane, which induces 
the fifth term of (\ref{effectsonxdot}). But this effect is only important for 
main-sequence star companions \cite{kbm+96,wex98} or rapidly rotating WD 
companions \cite{klm+00}, and can be neglected here.

The last term of (\ref{effectsonxdot}) is only valid if there is a
nearby third companion, that perturbs the orbit significantly, like in
the PSR~B1620$-$26 system \cite{ajrt96}. This is not the case for
PSR~J1012+5307 \cite{lwj+09}, as this would be well seen in the presence of higher oder derivatives of the rotational frequency of the pulsar \cite{jr97}. 

In summary, only the proper motion term~(\ref{xdotpm}) is important in our 
studies here. After accounting for this effect on $\dot{x}^{\rm obs}$ for every 
given $\Omega$, we calculate the contour plots of  $\hat{\alpha}_2$, and present 
them in the lower panels of figure~\ref{fig:a2_J1012}, for $i < 90^\circ$ (left) 
and $i > 90^\circ$ (right), respectively. In the calculation, $10^5$ Monte Carlo 
simulations are implemented to account for the measurement uncertainties of 
$\mu_\alpha$, $\mu_\delta$, $d$, $v_r$, $q$, $m_c$, and $\dot{x}$. As we can 
see, $\hat{\alpha}_2$ can be constrained to the order of $10^{-4}$ for most 
$\Omega$ realization. The $\Omega$ values in the figure where $\hat{\alpha}_2$ 
is virtually unconstrained correspond to the configurations when the angle 
between $\mathbf{w}$ and the orbital angular momentum $\psi \simeq 90^\circ$, 
or the angle between the projected $\mathbf{w}$ onto the orbital plane 
$\mathbf{w}_\perp$ and the direction of the ascending node $\vartheta
\simeq 90^\circ$ (see horizontal lines in the upper panels of
figure~\ref{fig:a2_J1012} for corresponding angles and compare them with the
divergencies in the lower panels). The reason for the divergencies is easy to 
see from (\ref{a2xdot}), where the right hand side vanishes when $\psi = 90^
\circ$, or $\vartheta = 90^\circ$, independent of $\hat{\alpha}_2$. In this 
situation $\hat{\alpha}_2$ cannot be constrained.


\subsection{PSR~J1738+0333}


\begin{figure}
\begin{center}
\includegraphics[width=14cm]{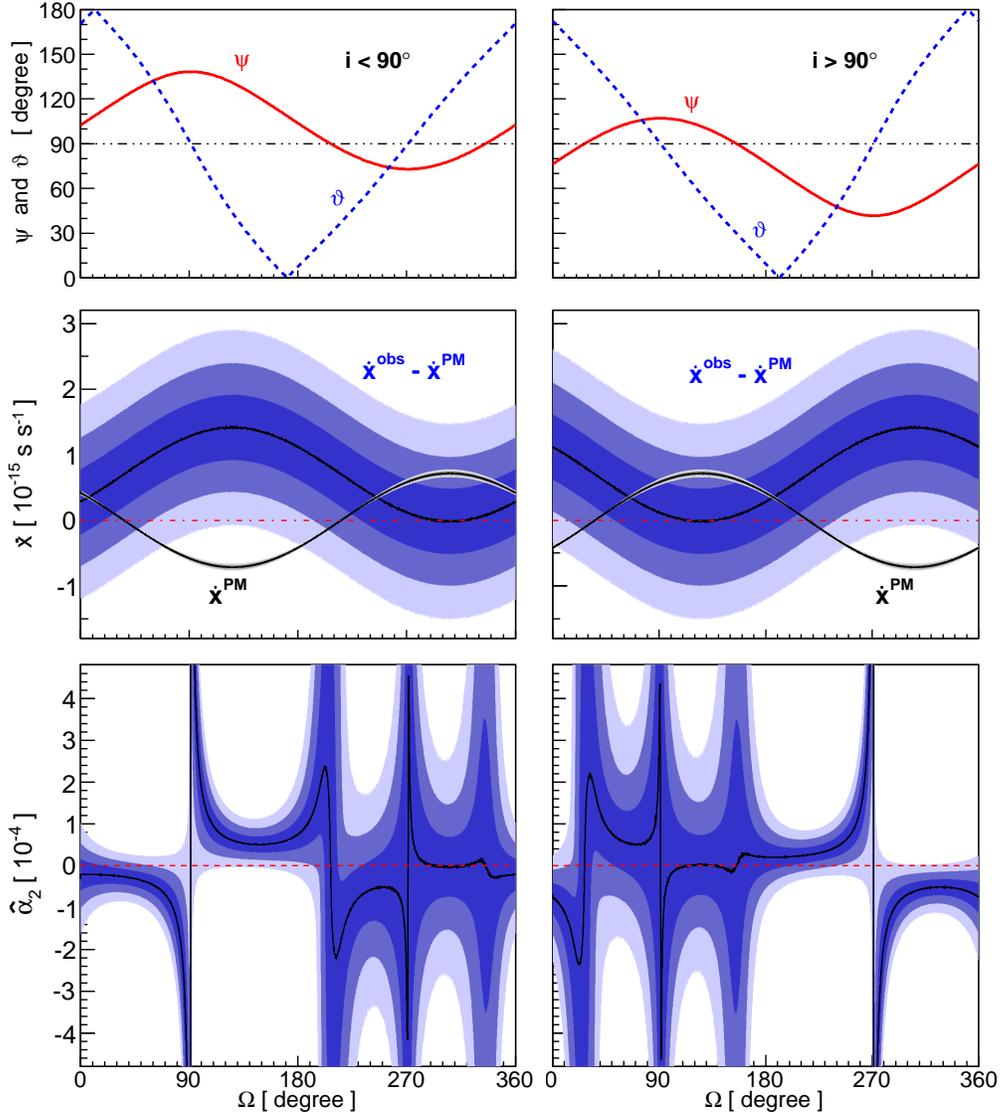}
\end{center}
\caption{\label{fig:a2_J1738}
  Same as figure~\ref{fig:a2_J1012}, for PSR~J1738+0333.}
\end{figure}


PSR~J1738+0333 is a small-eccentricity NS-WD binary system, with an orbital 
period of $\sim 8.5$\,h, which, together with other well measured physical 
quantities, makes it a superb astrophysical laboratory to test gravitational 
theories \cite{fwe+12}. The pulsar was discovered in 2001 in the Parkes high 
Galactic latitude survey \cite{jac05}, and later regularly timed with the 305-m
Arecibo telescope \cite{fwe+12}. It is one of the four millisecond pulsars known 
to be orbited by a WD companion bright enough for detailed spectroscopy 
\cite{akk+12}, among which, PSR~J1738+0333 is the most relativistic. Detailed 
optical studies of the WD companion and radio timing studies of the pulsar are 
presented in \cite{akk+12} and \cite{fwe+12}, respectively. Thanks to their 
studies, accurate binary parameters and spatial motion (transverse and radial) 
are available for the PSR~J1738+0333 binary system. Therefore, it 
also presents a good laboratory to study PFEs.

The strategy to constrain $\hat{\alpha}_2$ is the same as in the
case of PSR~J1012+5307. First, we get $\mathbf{w}$ from
$\mathbf{v}_{\mathrm{PSR-SSB}}$ and $\mathbf{v}_{\mathrm{SSB-CMB}}$,
and then the configuration of the system with respect to the CMB frame
can be obtained, as a function of $\Omega$, with a sign ambiguity of
$i$. Two important angles, $\psi$ and $\vartheta$, are depicted in the
upper panels of figure~\ref{fig:a2_J1738}, for $i<90^\circ$ (left) and
$i>90^\circ$ (right), respectively.

Along the line of arguments for PSR~1012+5307 one finds, using the results
of \cite{fwe+12} (see table~\ref{tab:pars} for binary parameters), that
also for PSR~J1738+0333 the only relevant term in
(\ref{effectsonxdot}) is the contribution by the proper motion of the
system, $\dot{x}^{\rm PM}$. Because of the more precise measurements
of the PSR~J1738+0333 parameters, the uncertainty of $\dot{x}^{\rm PM}$
is smaller correspondingly, as illustrated in the middle panels of
figure~\ref{fig:a2_J1738} for $i < 90^\circ$ (left) and $i > 90^\circ$
(right), respectively. The corresponding residual values for $\dot{x}$, after 
subtracting $\dot{x}^{\rm PM}$, are also depicted.

If we adopt the assumption that GR is the correct theory of gravity
for the PSR~J1738+0333 system, then we can get constraints on the
longitude of the ascending node $\Omega$. At a 95\% confidence level
one finds
\begin{equation}\label{eq:OmLT90}
\begin{array}{lcll}
  \Omega &\in& (0^\circ, 60^\circ) ~\mbox{or}~ (190^\circ,360^\circ) &
  \mbox{when $i < 90^\circ$} \,,\\
  \Omega &\in& (10^\circ, 240^\circ) &  
  \mbox{when $i > 90^\circ$} \,.
\end{array}
\end{equation}
We will partly use these results in the next section, where we constrain 
$\hat{\alpha}_1$. Naturally, to constrain $\hat{\alpha}_2$ we cannot use 
(\ref{eq:OmLT90}) as it is based on GR, i.e.~$\hat\alpha_2 \equiv 0$. The same 
is true for PSR~1012+5307, where the constraint of $\Omega$ can be found in 
(11--14) of \cite{lwj+09}.

In the calculation, for every $\Omega$ $10^5$ Monte Carlo simulations are 
implemented to account for the measurement uncertainties of $\mu_\alpha$,
$\mu_\delta$, $d$, $v_r$, $q$, $m_c$, and $\dot{x}$. The results are
plotted in the lower panels of figure~\ref{fig:a2_J1738} for $i <
90^\circ$ (left) and $i > 90^\circ$ (right), respectively. Similarly,
divergencies are caused by the unfavorable configurations with $\psi \simeq
90^\circ$ or $\vartheta\simeq 90^\circ$ (see upper panels for
reference), where $\hat{\alpha}_2$ can hardly be constrained. For
most $\Omega$, $\hat{\alpha}_2$ is constrained to be of order $
\sim5\times 10^{-5}$, about two times better than that of
PSR~J1012+5307.


\subsection{Probability distribution of $\hat{\alpha}_2$}


\begin{figure}
\begin{center}
\includegraphics[width=10.2cm]{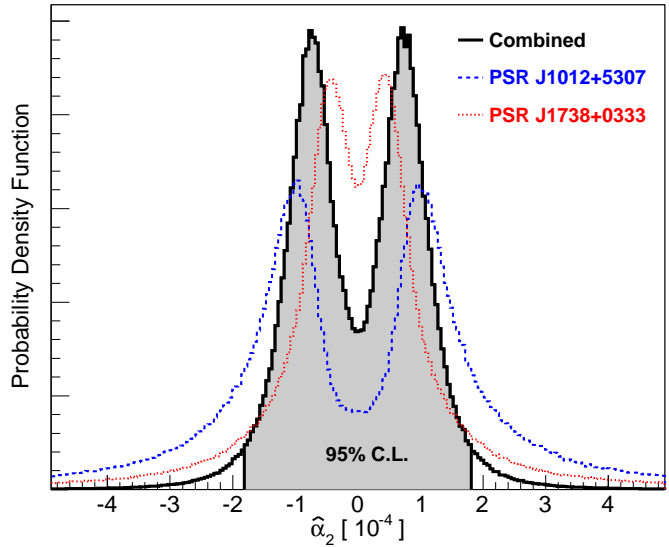}
\end{center}
\caption{Probability distributions of $\hat{\alpha}_2$ from
  PSR~J1012+5307 (blue dashed histogram), PSR~J1738+0333 (red dotted
  histogram), and their combination (black solid histogram). At 95\%
  confidence level, $|\hat{\alpha}_2|$ is constrained to be less than
  $1.8 \times 10^{-4}$ from the combined probability distribution.}
  \label{fig:a2_pdf} 
\end{figure}


We cannot directly constrain $\hat\alpha_2$ from (\ref{a2xdot}), and have to set 
up Monte Carlo simulations to account for our ignorance in $\Omega$ and 
ambiguity between $i$ and $180^\circ-i$. In the simulation, we assume that 
$\Omega$ is uniformly distributed between $0^\circ$ and $360^\circ$, and an 
equal probability to have $i < 90^\circ$ or $i > 90^\circ$. The measurement 
uncertainties of the spatial binary motions, the time derivative of the 
projected semi-major axis $ \dot{x}$, the binary mass ratio $q$, the companion
mass $m_c$, and also the distance from parallax measurement, are
properly accounted for, assuming a Gaussian error distribution.  
One million simulations for each of the two pulsars have been conducted. 
In every realization, we subtract the secular effects on $\dot{x}$ from proper 
motions, and calculate $\hat{\alpha}_2$ according to (\ref{a2xdot}).

The simulated results are summarized as probability distributions of
$\hat{\alpha}_2$ for PSRs~J1012+5307 and J1738+0333, illustrated in
figure~\ref{fig:a2_pdf} using a blue dashed histogram and a red dotted
histogram, respectively. The long tails of the probability
distributions from these two binaries are due to the undesirable
configurations when $\psi \simeq 90^\circ$ or $\vartheta \simeq
90^\circ$, which cause the divergencies in the lower panels of
figures~\ref{fig:a2_J1012}~and~\ref{fig:a2_J1738}. From these probability
distributions we find at 95\% confidence level,
\begin{eqnarray}
  |\hat{\alpha}_2| &<& 3.6\times10^{-4} \quad \mbox{for
    PSR~J1012+5307}\,,\\
  |\hat{\alpha}_2| &<& 2.9\times10^{-4} \quad \mbox{for
    PSR~J1738+0333}\,.
\end{eqnarray}

In figure~\ref{fig:a2_pdf}, we also show the probability distribution of
$\hat{\alpha}_2$ from the combination of these two pulsar binaries
(black solid histogram), assuming that their measurements are
independent, and that $\hat{\alpha}_2$ has only a weak functional
dependence on the NS mass in the range of $1.3$ -- $
2.0~{\rm M}_\odot$.  The combined probability distribution
demonstrates a much shorter and suppressed tail, which means it is
very unlikely that both systems are in the unfavorable configurations.
From the combined probability distribution, we obtain a constraint of
\begin{equation}\label{oura2}
|\hat{\alpha}_2| < 1.8 \times 10^{-4} \quad \mbox{(95\% C.L.)} \,.
\end{equation}
It is by three orders of magnitude better than the result of~\cite{wk07}, 
i.e.~(\ref{limita2}), although one has to keep in mind that the double pulsar 
tests the gravitational interaction of {\em two} strongly self-gravitating 
objects. This limit is still by three orders of magnitude weaker than the Solar 
system limit~(\ref{solara2}), but accounts for possible strong-field deviations 
in NSs. Compared with the Solar system limit from LLR~\cite{mwt08}, 
(\ref{oura2}) is still 3.6 times weaker. Being related to a secular effect, the 
limit on $\hat{\alpha}_2$ will improve fast with observing time $T_{\rm obs}$, 
namely proportional to $T_{\rm obs}^{-3/2}$. A disadvantage of this test, as 
compared for instance to the LLR experiment, is its dependence on probabilistic 
considerations with respect to the unknown angle $\Omega$. 


\section{A robust method to constrain $\hat{\alpha}_1$}
\label{sec:a1}

\begin{figure}
\begin{center}
\includegraphics[width=14cm]{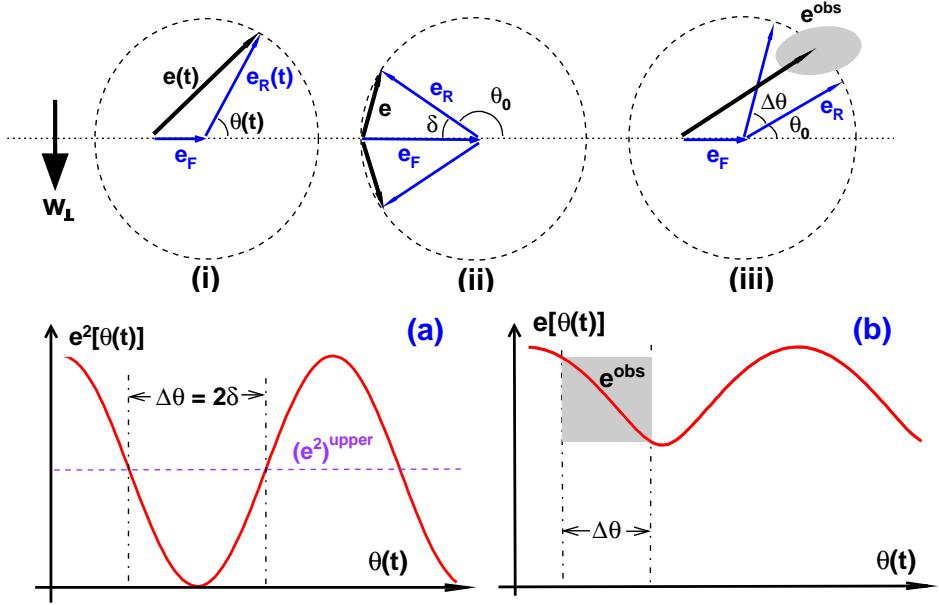}
\end{center}
\caption{Illustration of our robust method to constrain
  $\hat{\alpha}_1$.  {\it Upper:} (i) observational eccentricity
  $\mathbf{e}(t)$, is a vectorial superposition of a ``rotating
  eccentricity'' $\mathbf{e}_R(t)$ and a ``forced eccentricity''
  $\mathbf{e}_F$ \cite{de92a}; (ii) the worst starting configuration,
  where $\mathbf{e}_R(t)$ rotates $\Delta\theta = 2\delta$ during the
  observational time, and in the midtime, $\mathbf{e}_R(t)$ and
  $\mathbf{e}_F$ cancel out completely; this 
  applies to binaries where no explicit eccentricity is detected, like
  in PSR J1012+5307, and it constrains $\hat{\alpha}_1$ most
  conservatively; (iii) during the observational time,
  $\mathbf{e}_R(t)$ rotates out an angle $\Delta\theta$, and the
  time-averaged $\mathbf{e}$ and its variance should be consistent
  with measurements (see text for two criteria); this
  applies to binaries where an explicit
  eccentricity vector is detected, like in PSR J1738+0333.  {\it
    Lower:} (a) the sinusoidal evolution of $e^2(t)$ as a function of
  $\theta(t) = \theta_0 + \dot{\omega}_{\rm PN} t$; the indicated
  $\Delta\theta$ corresponds to the one of the 
  upper panel (ii); (b) the (not sinusoidal) time evolution of $e(t)$
  as a function of $\theta(t)$, and the indicated $\Delta\theta$
  corresponds to the one of the upper panel (iii).
  \label{fig:ga1}}
\end{figure}

As mentioned in section~\ref{sec:analytical}, Damour and Esposito-Far{\`e}se 
used a novel geometrical way to constrain $\hat{\alpha}_1$ with 
small-eccentricity binary pulsars~\cite{de92a}. In their paper they showed that 
the observed eccentricity vector, $\mathbf{e}(t)$, is a vectorial superposition 
of a ``rotating eccentricity'' $\mathbf{e}_R(t)$ (with constant length), and a ``forced eccentricity'' $\mathbf{e}_F$,
\begin{equation}\label{er+ef}
  \mathbf{e}(t) = \mathbf{e}_F + \mathbf{e}_R(t) \,,
\end{equation}
where $\mathbf{e}_R(t)$ is a vector of (unknown) constant magnitude which 
rotates in the orbital plane with angular velocity $\dot{\omega}_{\rm PN}$, and 
$\mathbf{e}_F$ is a fixed vector,
\begin{equation}\label{ef}
  \mathbf{e}_F = 
    \frac{\hat{\alpha}_1}{4 c^2} \, \frac{q-1}{q+1} \,
    \frac{n_b}{\dot{\omega}_{\rm PN}} \,  \mathcal{V}_O \,
    {\mathbf{k}} \times \mathbf{w} \,.
\end{equation}
A graphical illustration of this dynamics is given in the upper panel (i) of 
figure~\ref{fig:ga1}.


In analogy with the equivalence principle violation test in \cite{ds91}, a 
probabilistic reasoning is used in \cite{de92a} to constrain $\hat{\alpha}_1$. 
Because of two unknown angles, i.e., the time-dependent angle $\theta(t)$ 
between $\mathbf{e}_F$ and $\mathbf{e}_R(t)$, and the unknown angle of the 
longitude of the ascending node, $\Omega$, the probabilistic method has to 
assume a uniform distribution of $\theta(t)$ and $\Omega$ between $0^\circ$ and 
$360^\circ$. To 
make the ``random $\theta(t)$'' argument plausible, the binaries used in the 
test should be old enough to let $\mathbf{e}_R(t)$ have rotated several cycles
\cite{de92a,ds91}. Above two restrictions can be dropped in our robust method 
below.

In order to address a potential selection effect in the test of \cite{de92a}, 
when picking the system with the best figure of merit from a whole population of 
binary pulsars, an $\hat\alpha_1$ test has been developed in \cite{wex00} that 
extends the method of \cite{de92a} to the full related population, including 
those systems that have a low figure of merit. However, from the viewpoint of 
alternative gravity theories, the $\hat{\alpha}_1$ parameter might depend on the 
specific masses of the binary components used in the analysis. Such a dependence 
is well known for the generalized PPN parameters $\hat\gamma$ and $\hat\beta$  
\cite{wil93}. But in an analysis that 
combines different binary pulsar systems, one is forced to {\it a priori} 
assume a weak dependence of $\hat{\alpha}_1$ on the NS mass. Our new robust 
method can also naturally overcome the issues concerning the selection effect 
and the mass dependence, as it is based on a direct test of secular changes 
caused by $\hat{\alpha}_1$ in individual systems.

The basic idea behind our robust method is the expected change in the 
eccentricity vector during the observational span $T_{\rm obs}$, in case of a 
non-vanishing $\hat{\alpha}_1$. Because of the considerable periastron advance 
(see table~\ref{tab:pars} for GR values), and the fact that, by now, the timing 
observations span more than 10 years for PSRs~J1012+5307 and J1738+0333, the 
``rotating eccentricity'' $\mathbf{e}_R(t)$ has already swept out a sizable 
angle $\Delta\theta$ ($\geq10^\circ$ for both PSRs~J1012+5307 and J1738+0333). 
In the presence of a large $\mathbf{e}_F$, this would induce an observable 
change of the eccentricity vector (see figure \ref{fig:ga1}). We can use the 
tight constraints on the very small eccentricities of these two systems to limit 
such a variation of the eccentricity vector, and use this to directly constrain 
$\hat{\alpha}_1$ with remarkable precision. 

We will distinguish between two cases, depending on whether an upper 
bound or a positive measurement of the eccentricity vector is made from 
observations. If the eccentricity is not measured (like for PSR~J1012+5307), we 
use the measured (small) limit on the eccentricity to constrain $e_F$, and 
consequently $\hat\alpha_1$. If a positive detection of the eccentricity vector 
is made (like for PSR~J1738+0333), we include its directional 
information along with the smallness of its variation to constrain 
$\hat{\alpha}_1$.


\subsection{PSR~J1012+5307, a short orbital period system with an unmeasured 
eccentricity}
\label{sec:a1_1012}

For PSR~J1012+5307, at present there is no measurement of an orbital 
eccentricity, but a tight upper limit of the order of $10^{-7}$ (see table 
\ref{tab:pars}). This low limit, in combination with the fact that
the periastron should have precessed by about $10^\circ$ over the observing time 
span $T_{\rm obs}$, allows to put constraints on an $\hat\alpha_1$-related 
polarization of the orbit. In the following we will outline the method. 

From (\ref{er+ef}), we arrive at,
\begin{equation}\label{erefcomp}
  e^2 = e_F^2 + e_R^2 + 2 e_F e_R \cos \theta \,.
\end{equation}
Since $e_R$ and $e_F$ are both constant, the observational $e^2[\theta(t)]$ changes
as a sinusoidal function of $\theta(t)$, as shown in the lower panel (a) of 
figure~\ref{fig:ga1}. As $\theta(t) = \theta_0 + \dot{\omega}_{\rm PN} t$ is a 
linear function of time $t$, $e^2(t)$ is a sinusoidal function of $t$ as well. 
We also show a typical temporal evolution of $e[\theta(t)]$ in the lower panel 
(b) of figure~\ref{fig:ga1}, which is not sinusoidal.

For relativistic small-eccentricity binary systems, especially for those that 
are being observed for more then a decade, the vector $\mathbf{e}_R(t)$ has 
already swept out a non-negligible angle, which corresponds in the lower panels 
of figure~\ref{fig:ga1} to a span $\Delta\theta$ in the horizontal axis. This 
factor would induce an eccentricity variance because of the vectorial addition 
in (\ref{er+ef}) that scales with $\hat{\alpha}_1$. If then over a long time 
span an orbital eccentricity remains undetected to a small value, like for 
PSR~J1012+5307, one can directly constrain $|\hat{\alpha}_1|$ to some upper 
limit $|\hat{\alpha}_1|^{\rm upper}$, without any probabilistic assumptions 
about $\theta$. The reasoning is as follows. If $e_F \gg e_R$, then $\mathbf{e} 
\simeq \mathbf{e}_F$, hence the observational smallness of $e$ directly puts a 
limit on $e_F$. If $e_F \ll e_R$, then $\mathbf{e} \simeq \mathbf{e}_R$, hence 
$e_F \ll e_R \simeq e$. Therefore, a comparable magnitude of $e_F$ and $e_R$ 
which cancels them out (for an appropriate $\theta$)
would present the weakest constraint on $e_F$. But even 
in the unlikely event of complete cancellation, a sizable $ \hat{\alpha}_1$ 
cannot hide forever. A finite observational time $T_{\rm obs}$, hence a finite 
change in $\theta$, $\Delta \theta = \dot{\omega}_{\rm PN} T_{\rm obs}$, would 
induce a sizable variation of the eccentricity. We can use this ``induced'' 
eccentricity variation to perform the most conservative limit on 
$\hat{\alpha}_1$.

As we can see in the lower panel (a) of figure~\ref{fig:ga1}, for a given
observational time span, the most conservative configuration is the one
in which $\mathbf{e}_F$ and $\mathbf{e}_R(t)$ cancel out right at the middle
of the observational time span, as illustrated in the upper panel (ii). For the 
most conservative configuration, the ``rotating eccentricity'' sweeps out an 
angle $\theta$ from $\pi - \delta$ to $\pi + \delta$ during the 
observational time span $T_{\rm obs}$, where $\Delta \theta = 
\dot{\omega}_{\rm PN} T_{\rm obs} \equiv 2 \delta$. 

Before moving on, we would like to point out that our analysis accounts for the 
fact that the timing eccentricity published results from a fit to the whole 
observational data set of the pulsar binary, spanning 15 years. Hence the 
published eccentricity (table \ref{tab:pars}) represents a ``weighted'' average 
of a potentially changing eccentricity $e(t)$. After accounting for this, we can 
get an upper limit for the maximum eccentricity $\bar{e}$ hidden in the 
data, and from this a limit on $e_F$ from (\ref{erefcomp}),
\begin{equation}
  e_F \leq \frac{\bar{e}}{\sqrt{1-\sin^2\delta/\delta^2}} \,,
\end{equation}
which can be converted into an upper limit on $\hat{\alpha}_1$
through (\ref{ef}),
\begin{equation}\label{fig:limit_a1_1}
  |\hat{\alpha}_1|^{\rm upper} = 
    \frac{1}{\pi^2(q-1)} \,
    \frac{\bar{e} P_b^2}{x} \,
    \frac{\dot{\omega}_{\rm PN}}{\sqrt{1-\sin^2\delta/\delta^2}}\,
    \left(\frac{\sin i}{\sin\psi}\right) \,
    \left(\frac{w}{c}\right)^{-1} \,,
\end{equation}
where $\psi$ is again the angle between the orbital angular momentum
and $\mathbf{w}$ (see figure~\ref{geometry}).
On the right hand side of (\ref{fig:limit_a1_1}), for a given $\Omega$, all
quantities are observables or can be directly derived from
observational quantities, except $\dot{\omega}_{\rm PN}$. As given
in (\ref{omegaR}), $\dot{\omega}_{\rm PN}$ is the advance rate of
periastron, with potential corrections from the (generalized)
Eddington-Robertson-Schiff parameters and PFE parameters.   
The solution is to take
advantages of the smallness of $\delta$, as it is indeed the case in
the binaries which we will use to constrain $\hat{\alpha}_1$. For
PSRs J1012+5307 (and J1738+0333) the $\delta$ is smaller than
$10^\circ$. When $\delta$ is small, (\ref{fig:limit_a1_1}) becomes,
\begin{equation}\label{eq:limit_a1_2}
  |\hat{\alpha}_1|^{\rm upper} \simeq
    \frac{2\sqrt{3}}{\pi^2 (q - 1)} \,
    \frac{\bar{e} P_b^2}{x T_{\rm obs}} \,
    \left(\frac{\sin i}{\sin\psi}\right) \,
    \left(\frac{w}{c}\right)^{-1} \,,
\end{equation} 
where, to first order, $\dot{\omega}_{\rm PN}$ cancels out in the numerator and
denominator, since $\sqrt{1-\sin^2\delta/\delta^2} \approx \delta/\sqrt{3}
= \dot{\omega} T_{\rm obs} / (2 \sqrt{3})$. Strictly speaking, the smallness of 
$\delta$ for PSR~J1012+5307 (and PSR~J1738+0333) has been inferred from the GR 
value of $\dot\omega_{\rm PN}$. The argument would break down if there is a 
factor of a few deviation from GR in these systems. This we consider as 
unlikely, as there is neither such a deviation from GR in the gravitational wave 
damping of these systems, nor is such a large deviation seen in generic direct 
tests of $\dot\omega$, like in the double pulsar \cite{kw09}.


\begin{figure}
\begin{center}
\includegraphics[width=9.3cm]{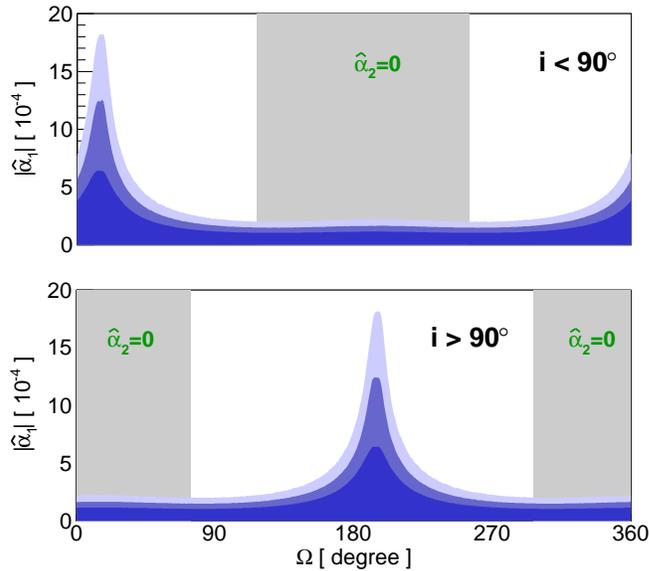}
\end{center}
\caption{Upper limit of $|\hat{\alpha}_1|$ as a function of the
  longitude of ascending node $\Omega$, derived from PSR~J1012+5307
  by using (\ref{eq:limit_a1_2}) ({\it Upper}: $i<90^\circ$; {\it
    Lower}: $i>90^\circ$). The shadowed regions are the allowed values
  of $\Omega$ at 95\% confidence level, assuming $\hat{\alpha}_2=0$
  \cite{lwj+09}. Different contours correspond to $68\%$,
  $95\%$, and $99.7\%$ confidence levels.\label{a1_J1012}}
\end{figure}


As mentioned in section~\ref{sec:a2}, PSR~J1012+5307 has been observed for
$15$ years (for the timing solution presented in
\cite{lwj+09}). During that time, its periastron has already rotated
out an angle $ \Delta \theta\simeq 10^\circ$, which corresponds to
$\delta\simeq5^\circ$ in the most conservative configuration. The
smallness assumption of $\delta$ is satisfied, hence we can use
(\ref{eq:limit_a1_2}) instead of the more rigorous expression
(\ref{fig:limit_a1_1}).

After properly accounting for all measurement errors, we carry out
$10^5$ Monte Carlo simulations to get an upper limit of
$\hat{\alpha}_1$ for every value of $\Omega$ between $0^\circ$ and $360^\circ$
(in steps of one degree). The results are depicted in
figure~\ref{a1_J1012} for $i<90^\circ$ (upper) and $i>90^\circ$
(lower), respectively. Our calculation uses the worst configuration
(see the upper panel (ii) of figure~\ref{fig:ga1}), hence the limit is
most conservative and reliable.  It is easily seen that the results
for $i<90^\circ$ and $i>90^\circ$ are merely shifted by
$180^\circ$. The existence of the peak near $15^\circ$ for the case
$i<90^\circ$ ($195^\circ$ for the case $i > 90^\circ$) is caused by
the $1/\sin\psi$ factor in (\ref{fig:limit_a1_1}). It can be understood
from the $\psi$ curve in the upper panels of figure~\ref{fig:a2_J1012}. It
is important that $\sin\psi$ does not vanish for 
$\Omega \in [0^\circ,360^\circ)$, as this avoids divergencies like in 
figure~\ref{fig:a2_J1012} for the $\hat{\alpha}_2$ test, which would have to 
be excluded based on probabilistic considerations.

From figure~\ref{a1_J1012}, most $\Omega$ realization would limit 
$|\hat{\alpha}_1|$ to be less than $2 \times 10^{-4}$ (95\% C.L.). Worth to 
mention that, our confidence level for the $\hat{\alpha}_1$ test is purely from 
measurement errors, in contrast with that from probabilistic assumptions. If we 
assume a random $\Omega \in [0^\circ,360^\circ)$, a similar constraint is 
obtained. However, since we want to constrain $\hat{\alpha}_1$ most robustly, we
conservatively adopt the worst configuration ($\Omega \simeq 15^\circ$ for 
$i < 90^\circ$ or $\Omega \simeq 195^\circ$ for $i > 90^\circ$) and get a limit
\begin{equation}\label{J1012a1noOmega}
  |\hat{\alpha}_1| < 1.3 \times 10^{-3} 
  \quad \mbox{(95\% C.L.)} \,.
\end{equation}
This limit is one order of magnitude worse than that obtained 
in \cite{wex00}, see (\ref{limita1}). But as it avoids the probabilistic 
considerations of the method used in \cite{wex00}, we consider this limit as 
more robust. Furthermore, this limit is more likely to improve in the future 
than that of \cite{wex00}, which we will discuss when giving the figure 
of merit of this test, in section \ref{subsec:fom}.

It is worth mentioning that, as we can see in
section~\ref{sec:analytical}, for small-eccentricity binaries, the
effects induced by $\hat{\alpha}_1$ and $\hat{\alpha}_2$
decouple. Hence this kind of test is not directly influenced by a
non-zero $\hat{\alpha}_2$. But a vanishing $\hat{\alpha}_2$ would
tighten the $\hat{\alpha}_1$ constraint a little further.  If we adopt
a zero $\hat{\alpha}_2$, or take the Solar limit (\ref{solara2}) for $\alpha_2$ 
as a limit for $\hat\alpha_2$, then the observed $\dot{x}$ of
\cite{lwj+09} can be attributed totally to the contribution of the proper
motion, i.e.\ (\ref{xdotpm}). Consequently, as mentioned, $\Omega$ can
be constrained to certain value ranges \cite{lwj+09}. We plot the consistent
$\Omega$ values (95\% C.L.) as shadowed regions in
figure~\ref{a1_J1012}. We can see that, this extra constraint excludes
the worst $\Omega$ configuration in both cases of $i<90^\circ$ and
$i>90^\circ$. Hence, with a vanishing $\hat{\alpha}_2$, we get a much
tighter (conservative) limit of,
\begin{equation}\label{eq:limit_a1_1012}
  |\hat{\alpha}_1| < 1.6 \times 10^{-4} 
  \quad \mbox{(95\% C.L.)} \,.
\end{equation}
This limit is comparable to the current best Solar system limit on $\alpha_1$, 
coming from LLR \cite{mwt08}, and is only slightly worse than 
the current best limit~(\ref{limita1}) for strongly self-gravitating bodies, but 
based on a method that avoids probabilistic considerations in terms of the exclusion of certain unfavorable angles.


\subsection{PSR~J1738+0333, a short orbital period system with a measured 
eccentricity}

\begin{figure}
\begin{center}
\includegraphics[width=9.3cm]{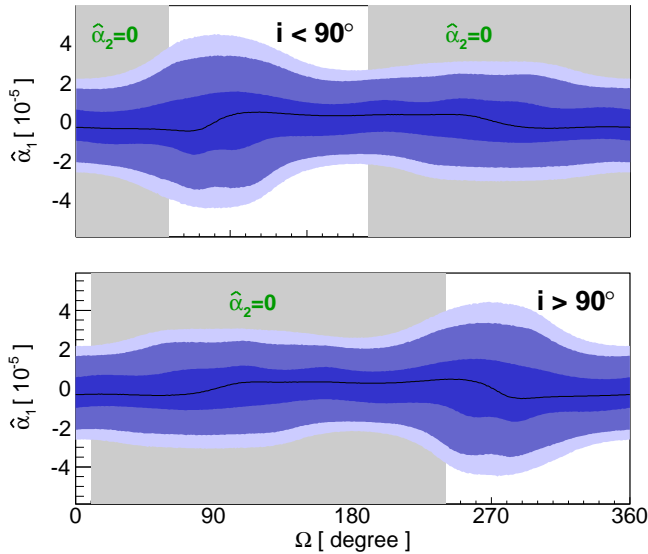}
\end{center}
\caption{\label{fig:a1_J1738} 
  The $\hat{\alpha}_1$ constraint from PSR~J1738+0333, as a function of the 
  longitude of ascending node $\Omega$ ({\it Upper}: $i<90^\circ$; {\it Lower}: 
  $i>90^\circ$). The limit is obtained from Monte Carlo simulations, by using 
  full information of the observed eccentricity vector. The shadowed regions are 
  the allowed values of $\Omega$ at 95\% confidence level, assuming 
  $\hat{\alpha}_2 = 0$, see (\ref{eq:OmLT90}). 
  Different contours correspond to $68\%$, $95\%$, and $99.7\%$ confidence 
  levels.}
\end{figure}

Like PSR~J1012+5307, PSR~J1738+0333 is also a short orbital period NS-WD
binary which can be use to constrain $\hat{\alpha}_1$. This binary
has been observed for 10 years \cite{fwe+12}, and during this time, the
periastron has advanced by $\Delta\theta \simeq 16^\circ$. Unlike
PSR~J1012+5307, PSR~J1738+0333 has a 3-$\sigma$ measurement of the (intrinsic)
orbital eccentricity vector (see $\eta$ and $\kappa$ in table~\ref{tab:pars}),
while any change in the eccentricity vector is still hidden in the measurement uncertainties. 
In a PFE test that exploits all the available information about PSR~J1738+0333,
this fact has to be taken into account (cf.~\cite{sfl+05}, where such a 
directional information, i.e.~the longitude of periastron $\omega$, has been 
used to improve the constraints on a violation of the strong equivalence 
principle). But let us first, for simplicity, make only use of the upper limit 
for the eccentricity (2-$\sigma$ upper limit for $e$: $5.7 \times 10^{-7}$) 
and apply the method discussed in the previous subsection on PSR~J1012+5307, as 
this method is fast and can be easily compared to the PSR~J1012+5307 result, 
which is based on the same method. For PSR~J1738+0333 one finds $\delta\simeq 
8^\circ$, and furthermore
\begin{equation}\label{1738a1}
  |\hat{\alpha}_1| < 1.6 \times 10^{-4} \quad 
  \mbox{(95\% C.L.)} \,,
\end{equation}
for the worst configuration of $\Omega$ and $\theta_0$ (i.e.\ upper panel (ii) 
of figure~\ref{fig:ga1}), without assuming a vanishing $\hat{\alpha}_2$. This 
limit has to be compared to (\ref{J1012a1noOmega}). The improvement over that 
of PSR~J1012+5307 comes from the shorter orbital period and a smaller 
$1/\sin\psi$ factor for the worst $\Omega$, as can be seen in the upper panels 
of figure~\ref{fig:a2_J1738}.

However, for PSR~J1738+0333 we have a $3\sigma$-measurement for the eccentricity 
vector, instead of just an upper limit on its magnitude (see 
table~\ref{tab:pars}). The result is expressed in terms of the first and
second Laplace-Lagrange parameters, $\eta \equiv e\sin\omega$ and
$\kappa\equiv e\cos\omega$~\cite{fwe+12}, from the {\tt ELL1} timing
model~\cite{lcw+01}. Therefore, essentially we have constraints on the
magnitude and the direction of the eccentricity vector. To fully include this   
information, we develop a new method, whose basic idea is depicted in the upper 
panel (iii) and the lower panel (b) of figure~\ref{fig:ga1}, and is to be 
elaborated below.

We \textit{a priori} have no knowledge about the magnitudes of $\mathbf{e}_F$ 
and $\mathbf{e}_R$, nor the initial angle between them, $\theta_0$, at the time 
when timing observation started. Instead, we have rough information about the 
superposed eccentricity $\mathbf{e}$, including its magnitude and direction, to 
$\sim3\sigma$ precision (see $\eta$ and $\kappa$ of PSR J1738+0333 in table 1), 
and also the direction of $\mathbf{e}_F$ as a function of the unknown $\Omega$. 
However, we cannot use $\mathbf{e}$ directly, for $\eta$ and $\kappa$ were 
treated as constants when fitting to TOAs, therefore they represent equivalently 
time-averaged quantities of the observational span.  We set up Monte Carlo 
simulations to select those $\hat{\alpha}_1$ which do not conflict with the 
timing observation of PSR J1738+0333.  Our simulation mainly includes the 
picking of an $\hat{\alpha}_1$ (hence $\mathbf{e}_F$) and $\mathbf{e}_R$ (a 
sufficiently large range for $e_R>0$ and a $\theta_0\in[0^\circ,360^\circ)$), 
for every $\Omega \in[0^\circ,360^\circ)$. 
The worst constraint of $\hat{\alpha}_1$ is announced as our most conservative 
limit. All measurement errors are properly considered in our simulations.

The practical implementation to get the constraints for $\hat{\alpha}_1$ is as 
following. First, we randomly choose $\hat{\alpha}_1$, $e_R$ ($>0$), uniformly 
from wide ranges, and $\theta_0$ from $[0^\circ,360^\circ)$, in order to map out 
a data-cube for these three values. Second, for every point in the data cube we 
let the eccentricity vector, $\mathbf{e}(t) = \mathbf{e}_F + \mathbf{e}_R(t)$, 
evolve under the dynamics given by the Lagrangian for a time span of 
observations. Then we calculate the time-averaged $\eta$ and $\kappa$, as well 
as their variances, $\sigma_\eta$ and $\sigma_\kappa$. We impose the following 
two criteria to get possible $\hat{\alpha}_1$, $e_R$, $\theta_0$ combinations 
which can still hide in the timing data: 1) The time-averaged $\eta$ and 
$\kappa$ should agree with observational values within $2\sigma$ measurement 
errors (we check that the results are practically unchanged if we use $1\sigma$ 
or $3\sigma$ instead); 2) The intrinsic variances of these two parameters, 
induced by their time evolution, should be smaller than the observational 
errors, because otherwise, they would contradict the actual measurement of 
$\eta$ and $\kappa$. By this, we assume that the reported measurement errors are 
the squared addition of the intrinsic variances and other possible errors, 
including those possibly from measurement devices and timing models. For each 
$\Omega$, we accumulate $10^5$ events in total, and get a distribution from it. 
The median values and distribution widths of $\hat{\alpha}_1$ are reported in 
figure~\ref{fig:a1_J1738} for $i < 90^\circ$ (upper) and $i > 90^\circ$ (lower), 
as a function of $\Omega$.

We can see from figure~\ref{fig:a1_J1738} that, $\hat{\alpha}_1$ is constrained 
to the level of $\sim 10^{-5}$. More importantly, in contrast to PSR~J1012+5307, 
the limit only weakly depends on the (presently) unknown angle of the ascending 
node, $\Omega$. For the worst configurations ($\Omega \simeq 92^\circ$ for 
$i < 90^\circ$ and $\Omega \simeq 273^\circ$ for $i > 90^\circ$) one finds
\begin{equation} \label{eq:a1_limit_1738}
  \hat{\alpha}_1 = -0.4^{+3.7}_{-3.1} \times 10^{-5} \quad 
                   \mbox{(95\% C.L.)} \,,
\end{equation}
which is about 40 times stronger than the limit from PSR~J1012+5307 
(cf.~(\ref{J1012a1noOmega})). As outlined above, this limit is free of any 
probabilistic considerations related to unknown angles. 
The limit in (\ref{eq:a1_limit_1738}) is more than five times better
than the present best limit on $\alpha_1$, coming from LLR \cite{mwt08}. 
Furthermore, it is also about four times better than the less robust test 
of \cite{wex00}.

Like in the case of 
PSR~J1012+5307, we could adopt the very small Solar system limit for $\alpha_2$
as a limit for $\hat{\alpha}_2$, in order to constrain the range of $\Omega$ 
with the help of $\dot{x}$. But here this would only slightly improve compared 
to the above constraint, because of the weak dependence on $\Omega$. 

It is also instructive to extract from our calculations the possible values of 
$e_F$, in comparison to the observed eccentricity $e$. We find in our Monte 
Carlo simulations that $e_F < 1.4 \times 10^{-6}$ (95\% C.L.) for the most 
conservative configuration in terms of $i$ and $\Omega$. Compared to the 
(Shapiro corrected) observed $e = \sqrt{\eta^2 + \kappa^2} = 3.4 \times 
10^{-7}$, this is only a factor of a few larger. This fact reveals a generic 
feature of our test, that a too large $e_F$ cannot hide in the timing data. It 
lays down the reasoning of the $\hat{\alpha}_1$ test presented here, utilizing 
long-term pulsar timing of small-eccentricity binaries.

The method developed here for PSR~J1738+0333, is capable of incorporating the 
directional information of a measured eccentricity vector to constrain 
$\hat{\alpha}_1$. In principle it can also be applied to binary pulsars where no 
eccentricity is measured, and the direction of the eccentricity vector is 
unconstrained. For instance, for PSR~J1012+5307 this method yields  
\begin{equation}
  \hat{\alpha}_1 = -0.0^{+1.1}_{-1.0}\times10^{-3} 
  \quad \mbox{(95\% C.L.)}. 
\end{equation}
As expected, this limit is comparable with that of (\ref{J1012a1noOmega}), and 
constitutes a nice test of the implementation used PSR~J1738+0333. In general, 
for systems that do not have a measured eccentricity, the method outline in 
section~\ref{sec:a1_1012} is preferable, as the method of this subsection is 
computationally considerably more expensive.


\section{Discussions and summary}
\label{sec:sum}

\subsection{Constraints on PFEs from the Galactic frames}

When using the isotropic CMB frame as the assumed preferred frame, we are 
basically assuming that the preferred frame is determined by the global matter
distribution in the Universe, and that the extra vectorial or tensorial 
components of gravitational interaction are long range, at least comparable to 
the Hubble radius. While this is generally the most plausible assumption, it is
still interesting to consider other, more local preferred frames, like the one
related to the rest frame of our Galaxy, or a frame in Galactic co-rotation with 
the overall local matter at the pulsar's location (cf.\ \cite{skmb08}). It is 
straightforward to apply the computations of this paper to these two Galactic 
frames.

We used the Galactic model of \cite{pac90}, which assumes a distance
of 8.0 kpc between the Solar system and the Galactic center, to
extract the Galactic rotation curve. Then the velocities of the pulsar
binary with respect to the Galactic frame and the local co-rotating
Galactic frame are obtained. They are typically smaller than the
velocity with respect to the CMB frame (see
table~\ref{tab:limits}). The binary velocity with respect to the local
co-rotating frame is about $100\,{\rm km\,s}^{-1}$, which is, however,
one order of magnitude larger than that of the Solar system
\cite{mig00}. Hence, pulsar binaries have the advantage to probe
secular PFEs with respect to locally co-rotating Galactic frames,
thanks to their peculiar velocity produced by the supernova, while
Solar system tests are expected to be clearly less sensitive to such
PFEs, because of the small peculiar Solar
velocity.\footnote{Nevertheless, see \cite{skmb08} for a constraint of
  $\alpha_1 = (1.6\pm8.0) \times 10^{-3}$ (95\% C.L.) from LLR when
  choosing the Barycentric Celestial Reference System as the preferred
  frame, based on periodic effects.} Table~\ref{tab:limits}
gives, besides the limits for the CMB frame, the limits for the two
Galactic frames introduced above, based on the methods presented in
sections~\ref{sec:a2} and \ref{sec:a1}. The limits concerning the two
Galactic frames are somewhat weaker than that of the CMB frame,
because of the smaller peculiar velocities of the pulsar
binaries. Nevertheless, they could be of interest for tests of
specific gravity theories that might predict or even require (to pass
Solar system tests) vector or tensor fields which are aligned with the
Galactic or local matter distribution.

\begin{figure}
\begin{center}
\includegraphics[width=12cm]{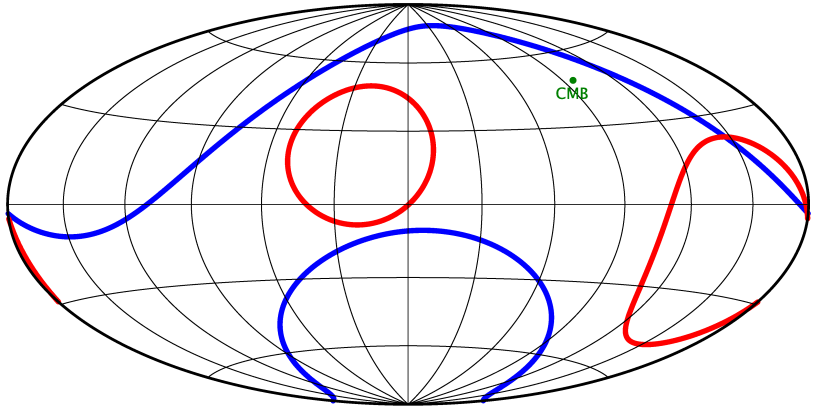}
\end{center}
\caption{\label{fig:pfesky}
  Sky coverage in the $\hat\alpha_1$ test by PSRs~J1012+5307 and J1738+0333.
  The blue curves mark the directions in the sky where PSR~J1012+5307 is 
  insensitive, and the red curves those directions where PSR~J1738+0333 is
  insensitive, due to the unknown $\Omega$.
  The sky is plotted in an Hammer-Aitoff projection using Galactic
  coordinates $l$ and $b$. The longitude $l$ increases from right to left, 
  from $l = -180^\circ$ to $l = +180^\circ$, while the latitude $b$ runs 
  from $b = -90^\circ$ to $b = 90^\circ$ from bottom to top. The grid gives 
  steps of $30^\circ$. The plot is based on a $w = 369\,{\rm km\,s^{-1}}$ for 
  the velocity of the SSB with respect to a potential preferred frame. The label 
  ``CMB'' denotes the direction of motion with respect to the CMB frame.
} 

\end{figure}

Finally, it is interesting to see the sky coverage of the two binary pulsars
in terms of a sensitivity towards a preferred frame. In the spirit of a ``PFE
pulsar antenna array'', proposed in \cite{wk07}, two suitable binary pulsars
can probe for a preferred reference frame in (almost) any direction in the sky.
Figure~\ref{fig:pfesky} shows the combined sky coverage in the $\hat\alpha_1$
test for PSRs~J1012+5307 and J1738+0333, for $w = 369 \, {\rm km\,s^{-1}}$,
the same magnitude as the velocity of the SSB with respect to the CMB. The 
figure nicely illustrates why PSR~J1738+0333 is more suitable to test the CMB 
frame, since the ``CMB'' dot is far away from any of the red curves, meaning 
that the CMB direction lies in the area where the system is particularly 
sensitive. The absence of any divergencies in figure~\ref{fig:a1_J1738} is a 
consequence of this.

\subsection{Strong-field modifications}

When comparing the results of this paper, obtained from pulsar-WD systems, with Solar system experiments, one has to keep in mind that alternative gravitational theories, in general, are expected to predict strong-field modifications of the PPN parameters due to the strong internal gravitational field of the pulsar. As
an example, in scalar-tensor gravity the PPN parameter $\gamma$ generalizes to
\begin{equation}
  \hat\gamma \equiv \gamma_{AB} = 
  1 - \frac{2\alpha_A\alpha_B}{1 + \alpha_A\alpha_B} \,,
\end{equation}
for a binary pulsar system, where $\alpha_A$ and $\alpha_B$ are the effective 
scalar coupling constants of pulsar and companion, respectively \cite{de92b}. 
The weak-field PPN parameter $\gamma$ is recovered for $\alpha_A = \alpha_B = 
\alpha_0$. In GR one has $\hat\gamma = \gamma = 1$. In the strong-field regime 
of a NS, $\hat\gamma$ can 
deviate significantly from $\gamma$ due to strong-field scalarization effects
\cite{de93}. Similarly, we may expect that $\hat\alpha_1$ and $\hat\alpha_2$
deviate from their PPN correspondents, $\alpha_1$ and $\alpha_2$. In the
absence of non-perturbative effects, one can illustrate this as an expansion
in the compactnesses $c_A$ and $c_B$ of the bodies \cite{de96}. In our case,
we would write something like 
\begin{equation}\label{expanse}
  \hat\alpha_a = \alpha_a + {\cal K}_i c_i + {\cal K}_{ij} c_i c_j + \dots \,,
\end{equation}
where $a=1,2$ and ${\cal K}_i$ and ${\cal K}_{ij}$ are coefficients 
characterizing deviations from general relativity, and $c_i \sim Gm_i/R_ic^2$ 
with mass $m_i$ and radius $R_i$ of body $i$. The compactnesses 
for the Earth and the Sun are roughly $c_\oplus \sim 10^{-10}$ and $c_\odot \sim 
10^{-6}$, respectively, which suppress ${\cal K}_i$- and ${\cal K}_{ij}$-related 
physical effects dramatically. In contrast, NSs have $c_{\rm NS} \sim 0.2$, 
which is one of the reasons why pulsar timing experiments are ideal probes for 
gravity effects associated with strong gravitational fields. Consequently, in a 
NS-WD system ($c_{\rm WD} \sim 10^{-4}$) we could still have a significant  
$\hat\alpha_a$, even if there is a tight Solar system constraint for $\alpha_a$.
This, for instance, supports the importance of the $\hat\alpha_2$ limits 
obtained in this paper.

Finally, when discussing the constraints on parameters of alternative gravity 
theories, one should be aware of a potential compactness-dependent (or mass-
dependent) nature of these parameters. Especially when combining different 
systems, like the $\hat{\alpha}_1$ test in \cite{wex00}, or the combined 
probability distribution function of the $\hat{\alpha}_2$ test presented in this 
paper (black solid histogram of figure~\ref{fig:a2_pdf}). Such tests implicitly 
assume that the parameter is approximately the same for all systems under 
investigation. We are aware of  this potential problem in our calculation. In 
the case of the PSRs J1012+5307 and  J1738+0333 binary systems one can argue 
that the similarity in the masses justifies such an assumption. However, in the 
presence of phenomena related to  some critical mass, like the spontaneous 
scalarization discovered in \cite{de93}, even a small difference in masses does 
not allow such an  assumption.  Our proposed robust test for $\hat{\alpha}_1$ 
overcomes this ``mass-dependence'' problem by only considering one system. Hence 
our final results on $\hat{\alpha}_1$ are more suitable to be quoted along with 
mentioning the specific system were it was obtained from and the related 
neutron-star mass.


\subsection{Figures of merit and further potential improvements}
\label{subsec:fom}

Finally, let us discuss potential improvements of the current limits of 
$\hat{\alpha}_1$ and $\hat{\alpha}_2$. For this, one conveniently identifies 
the figures of merit for the different tests. Besides details related to the 
shape and the size of the orbit, various geometrical angles and the barycentric 
velocity with respect to the preferred frame ($\mathbf{w}$) play a role in our 
tests. Therefore, in principle our figures of merit would depend on the 
geometrical configuration of binary systems under consideration, as well as 
their proper motions with respect to the preferred frame. However, after we drop 
the geometrical dependencies and ignore potential difference in 
$\mathbf{w}$, we can roughly get a figure of merit of our tests.

As for the $\hat{\alpha}_1$ test, the traditional method has a figure of merit, 
$P_b^{1/3}/e$ \cite{de92a}, which means that the strength of this test is not to 
improve until new systems with higher $P_b^{1/3}/e$ are discovered. In contrast, 
our figure of merit for the robust $\hat{\alpha}_1$ test is $T_{\rm obs} /
(P_b^{4/3}\bar{e})$ for binary pulsars with unmeasured eccentricities 
$e < \bar{e}$ (cf.~(\ref{eq:limit_a1_2})). Besides the discoveries of new 
systems with smaller $P_b^{4/3} \bar{e}$, the constraint has 
the potential to improve when the observational span becomes longer. In fact, it 
improves as $T_{\rm obs}^{3/2}$, as long as $e$ remains smaller than $\bar{e}$, 
and $\dot{\omega} T_{\rm obs} \lesssim 1$. For a binary pulsar with measured 
eccentricity $e$, we get a similar figure, where $\bar{e}$ is to be replaced 
with $\sigma_e$, the measurement error of the eccentricity vector, or that of 
the first and second Laplace-Lagrange parameters, namely $\sigma_\eta$ and 
$\sigma_\kappa$. If in the future the secular evolution of the eccentricity 
vector in PSR~J1738+0333, due to the relativistic periastron advance, can be 
measured, we could further constrain a potential polarization of the orbit 
caused by a non-vanishing $\hat\alpha_1$, or even detect the presence of a 
significant PFE eccentricity ${\bf e}_F$.

In contrast to our secular effects, the limit from LLR is based on periodic 
effects (see e.g. \cite{dv96}), and therefore only improves as 
$T_{\rm obs}^{1/2}$. 
Moreover, to test PFEs in LLR one has to deal with the motion of the Earth-Moon 
system around the Sun, introducing tidal forces and an annually changing 
$\mathbf{w}$, hence leading to the theoretical complexity of a three-body 
problem in the presence of a preferred frame \cite{dv96}.

For the $\hat{\alpha}_2$ test one finds the figure of merit
to be $1/(P_b^{1/3} \sigma_{\dot{x}})$ from (\ref{a2xdot}), where 
$\sigma_{\dot{x}}$ is the measurement uncertainty of $\dot{x}$. Hence, more 
relativistic systems (smaller $P_b$) with high timing precision (especially of 
$\dot{x}$) are advantageous to do the $\hat{\alpha}_2$ test. For the systems
discussed here, namely PSRs J1012+5307 and J1738+0333, persistent 
timing observations will reduce the measurement uncertainty of 
$\sigma_{\dot{x}}$. Hence the $\hat{\alpha}_2$ test will improve continuously as 
$T_{\rm obs}^{3/2}$, in contrast with that of \cite{nor87}. An even faster improvement is expected to come from new receiver and backend technologies 
and new telescopes.

It is worth noting that a measurement of the unknown longitude of the ascending 
node, $\Omega$, would improve both the $\hat{\alpha}_1$ and the $\hat{\alpha}_2$ 
test. If $\Omega$ can be determined independently, even rough constraints on 
$\Omega$ would make these tests more efficient. It would eliminate the 
(systematical) ``double peak'' structure in the probability distribution 
function of $\hat{\alpha}_2$ in figure~\ref{fig:a2_pdf}. Also it would select a
specific limit on $\hat{\alpha}_1$ from figures~\ref{a1_J1012} or
\ref{fig:a1_J1738}, where presently we are conservatively using the worst
$\Omega$ configuration. Unfortunately, neither pulsar timing nor optical 
astrometry are likely to provide such a measurement in the near future. Maybe
scintillation measurements would be able to provide interesting constraints on
$\Omega$, like this is the case for the double pulsar (Rickett {\em et al}, 
in prep.). 


\subsection{A brief summary of the results}

In summary, we presented an extended orbital dynamics of pulsar binaries under 
the influence of preferred frame effects that accounts for both generalized PPN
parameters, $\hat{\alpha}_1$ and $\hat{\alpha}_2$. In the limit of a small 
eccentricity, orbital effects from $\hat{\alpha}_1$ and $\hat{\alpha}_2$ 
decouple. We implemented two new methods to constrain $\hat{\alpha}_1$ and 
$\hat{\alpha}_2$, by directly constraining secular orbital changes expected from 
a violation of local Lorentz invariance in the gravitational sector. Both 
methods have been applied to the two binary pulsars PSRs J1012+5307 and 
J1738+0333, where we have full spatial velocity information. For a frame at rest 
with respect to the CMB, the best limit we obtain is
\begin{equation}
  \hat{\alpha}_1 = -0.4^{+3.7}_{-3.1} \times 10^{-5} 
  \quad \mbox{(95\% C.L.)} \,,
\end{equation}
which avoids the probabilistic considerations of previous methods, and clearly 
surpasses the current best limits obtained with both, weakly (Solar system) and 
strongly (binary pulsars) self-gravitating bodies, namely (\ref{llra1}) and 
(\ref{limita1}).

Concerning $\hat{\alpha}_2$, the best limit we obtain is
\begin{equation}
  |\hat{\alpha}_2| < 1.8 \times 10^{-4} \quad (95\%~\textrm{C.L.})\,.
\end{equation}
This limit is still three orders of magnitude weaker than the weak-field limit 
obtained in the Solar system, but constrains possible deviations related
to the strong internal gravitational fields of NSs. The limit
here surpasses the current best limit for strongly self-gravitating bodies, 
namely (\ref{limita2}), by three orders of magnitude, although strictly speaking
they are different in their physical nature, as (\ref{limita2}) probes the
interaction between {\em two} strongly self-gravitating bodies, in contrast to 
the pulsar-WD systems used in this paper. A drawback of the $\hat\alpha_2$ 
limit presented here is that it is still based on probabilistic considerations 
in excluding unfavorable values of the longitude of the ascending node $\Omega$,
and the combination of two systems with different NS masses. 

Our new methods promise continuous improvements with on-going timing 
observations of known systems, as well as the discovery of new suitable systems.
In addition, new receiver and backend technologies as well as new telescopes, 
like FAST \cite{nlj+11} and SKA \cite{sks+09}, will tremendously improve these 
tests.


\section*{Acknowledgments}

We thank John Antoniadis, William Coles, Paulo Freire, Michael Kramer and 
Clifford Will for stimulating discussions. We are grateful to Michael Kramer 
for carefully reading the manuscript. Lijing Shao is supported by China 
Scholarship Council (CSC).





\begin{landscape}
\begin{table}
\caption{Relevant parameters in the PFE calculations for
  PSRs~J1012+5307 \cite{lwj+09,lcw+01} and J1738+0333
  \cite{fwe+12,akk+12}.\label{tab:pars}}
  \begin{indented}
  \item[]\begin{tabular}{@{}lcc}
    \br
    Pulsars & PSR~J1012+5307 & PSR~J1738+0333 \\
    \mr
    Right Ascension, $\alpha$ (J2000) & ${\rm 10^h12^m33^s \hspace{-1.2mm}. 
      4341010(99)}$ & 
    ${\rm 17^h38^m53^s \hspace{-1.2mm}. 9658386(7)}$ \\
    Declination, $\delta$ (J2000) & $53^\circ07^\prime02^{\prime\prime} 
    \hspace{-1.2mm} .  60070(13)$ & 
    $03^\circ33^\prime10^{\prime\prime} 
    \hspace{-1.2mm} .  86667(3)$ \\
    Proper motion in $\alpha$, $\mu_\alpha~(\textrm{mas\,yr}^{-1})$ & 
    2.562(14) & 7.037(5) \\
    Proper motion in $\delta$, $\mu_\delta~(\textrm{mas\,yr}^{-1})$ & 
    $-$25.61(2) & 5.073(12) \\
    Distance, $d$ (kpc) & 0.836(80) & 1.47(10) \\
    Spin period, $P$ (ms) & 5.255749014115410(15) & 5.850095859775683(5) \\
    Orbital period, $P_b$ (d) & 0.60467271355(3) & 0.3547907398724(13) \\
    Projected semi-major axis, $x$ (lt-s) & 0.5818172(2) &
    0.343429130(17) \\
    $\eta \equiv e \sin\omega~(10^{-7})$$^{\rm a}$ & $-1.4\pm3.4$ & 
    $-1.4\pm1.1$ \\
    $\kappa \equiv e \cos\omega~(10^{-7})$ & $0.6\pm3.1$ & $3.1\pm1.1$ \\
    Time derivative of $x$, $\dot{x}~(10^{-15}~\textrm{s\,s}^{-1})$ & 2.3(8) & 
    0.7(5) \\
    Time derivative of $P_b$, $\dot{P}_b~(10^{-15}~\textrm{s\,s}^{-1})$
    & 50(14) & $-$17.0(3.1) \\ 
    Radial velocity, $v_r$ ($\textrm{km~s}^{-1}$) & 44(8) & $-$42(16)\\
    Mass ratio, $q \equiv m_p/m_c$ & 10.5(5) & 8.1(2) \\
    WD mass, $m_c~(\textrm{M}_\odot)$ & 0.16(2) & $0.181^{+0.008}_{-0.007}$ \\
    Pulsar mass, $m_p~(\textrm{M}_\odot)$ & 1.64(22) & $1.46^{+0.06}_{-0.05}$ \\
    Mass function, $f~(\textrm{M}_\odot)$ & 0.000578 & 0.0003455 \\
    Inclination, $i$ (deg) &  52(4) or 128(4) & 32.6(10) or 147.4(10) \\
    Advance of periastron (in GR), $\dot{\omega}_{\rm PN}^{(\rm GR)}
    ({\rm deg\,yr}^{-1})$ & 0.69(6) & 1.57(5) \\
    ``Characteristic'' velocity (in GR), $\mathcal{V}_O^{\rm (GR)}\
    ({\rm km\,s}^{-1})$ & 308(13) & 355(5) \\
    \br
  \end{tabular}
  \item[] $^{\rm a}$ Intrinsic $\eta$, after
      subtraction of the contribution 
      from the Shapiro delay according to (A22) of
      \cite{lcw+01}.
  \end{indented}
\end{table}
\end{landscape}


\begin{landscape}
\begin{table}
\caption{\label{tab:limits}
  Limits on the PFE parameters $\hat{\alpha}_1$ and $\hat{\alpha}_2$ at 95\% 
  confidence level, from the NS-WD binaries, 
  PSRs~J1012+5307 and J1738+0333 (see text for details).}
\begin{indented}
\item[] 
\begin{tabular}{llccc}
\br 
Pulsar Binary & Preferred Frame & $w$\,(km\,s$^{-1}$) &
$\hat{\alpha}_1$ & $\hat{\alpha}_2$ \\ \mr
%
%
\multirow{3}{*}{J1012+5307} & CMB & $477(14)$ & 
$|\hat{\alpha}_1| < 1.3 \times 10^{-3}$ &
$|\hat{\alpha}_2| < 3.6 \times 10^{-4}$ \\  
~ & Galaxy & $157(15)$ &
$|\hat{\alpha}_1| < 8.3 \times 10^{-3}$ &
$|\hat{\alpha}_2| < 7.9 \times 10^{-3}$ \\  
~ & Local Galactic rotation & $114(23)$ &
$|\hat{\alpha}_1| < 7.5 \times 10^{-3}$ &
$|\hat{\alpha}_2| < 1.1 \times 10^{-2}$ \\ \mr
%
%
\multirow{3}{*}{J1738+0333} & CMB & $327(6)$ & 
$\hat{\alpha}_1 = -0.4^{+3.7}_{-3.1} \times 10^{-5}$ &
$|\hat{\alpha}_2| < 2.9 \times 10^{-4}$ \\  
~ & Galaxy & $265(6)$ & 
$\hat{\alpha}_1 = -0.3^{+4.5}_{-4.0} \times 10^{-5}$ & 
$|\hat{\alpha}_2| < 8.3 \times 10^{-4}$ \\ 
~ & Local Galactic rotation & $82(10)$ &
$\hat{\alpha}_1 = +0.1^{+4.2}_{-4.1} \times 10^{-4}$  &
$|\hat{\alpha}_2| < 1.0 \times 10^{-2}$ \\ \mr
\multirow{3}{*}{Combined} & CMB & -- & -- &
$|\hat{\alpha}_2| < 1.8 \times 10^{-4}$ \\ 
~ & Galaxy & -- & -- &
$|\hat{\alpha}_2| < 4.5 \times 10^{-4}$ \\ 
~ & Local Galactic rotation & -- & -- &
$|\hat{\alpha}_2| < 3.4 \times 10^{-3}$ \\ \br
\end{tabular}
\end{indented}
\end{table}
\end{landscape}

\end{document}